\documentclass[11pt,a4paper]{article}
\usepackage{t1enc}
\usepackage[latin1]{inputenc}
\usepackage[english]{babel}
\usepackage[numbers]{natbib}
\usepackage{amssymb,stmaryrd}
\usepackage{wasysym}
\usepackage{graphicx}

\newcommand{\R}{\mathbb R}

\newcommand{\beq}{\begin{equation}}
\newcommand{\eeq}{\end{equation}}
\newcommand{\beqarr}{\begin{eqnarray}}
\newcommand{\eeqarr}{\end{eqnarray}}
\newcommand{\beqa}{\begin{eqnarray*}}
\newcommand{\eeqa}{\end{eqnarray*}}
\begin{document}
\thispagestyle{empty}

\title{\bf MOND-like acceleration in integrable Weyl geometric gravity}
\author{{\em Erhard Scholz}\footnote{University of Wuppertal, Department C, Mathematics, and Interdisciplinary Centre for History and Philosophy of Science; \quad  scholz@math.uni-wuppertal.de}}
\date{Oct. 15, 2015 } 
\maketitle
\begin{abstract}
In this paper a Weyl geometric scalar tensor theory of gravity with  scalar field $\phi$ and  scale invariant cubic (``aquadratic'')  kinetic Lagrangian is introduced. Einstein gauge (comparable to Einstein frame in Jordan-Brans-Dicke theory) is  most natural for studying trajectories. In it, the Weylian scale connection   induces an additional acceleration which in the weak field, static, low velocity limit  acquires the  deep MOND form of Milgrom/Bekenstein's gravity. The energy momentum of $\phi$ leads to another add on to Newton acceleration. Both additional accelerations together imply a MOND-ian phenomenology of the model. It has unusual   transition functions $\mu_w(x), \nu_w(y)$. They imply higher phantom energy density than in the case of the more common MOND models with transition functions $\mu_1(x), \, \mu_2(x)$. A considerable part of it is due to  the scalar field's energy density which, in our model,  gives a scale and generally covariant expression for  the self-energy of the gravitational field. 
\end{abstract}

{\small \tableofcontents}

\section{\small Introduction}
Shortly after  Milgrom originally  proposed his modified Newtonian dynamics, MOND, as an explanation for the observed anomalies in galaxy rotation curves, he and  Bekenstein showed how a MONDian dynamics could be derived from a Lagrangian of a scalar field $\phi$. It  involved a kinetic term of the scalar field, proportional to  $\tilde{f}(a_o^{-2}(\nabla \phi)^2)$ with a non-linear functional $\tilde{f}$ \cite{Bekenstein/Milgrom:1984}.\footnote{$a_o$ denoted the typical new constant of the MOND hypothesis, $a_o \approx \frac{1}{6} H_o\, [c]$, $H_o$ the Hubble constant in time units, $c$ the velocity of light.}
 A case distinction between the Newton regime and the MOND regime had to be  inbuilt into the functional $\tilde{f}$.
In the appendix of their paper they indicated how their  ``a-quadratic'' (AQUAL) Lagrangian could be adapted to  general relativity in a  Jordan-Brans-Dicke (JBD)  framework. This approach was the first of a collection of different attempts to cope with MOND phenomenology in  general relativistic frameworks (TeVeS, Einstein aether, and others). 
The relativistic a-quadratic Lagrangian approach itself (``RAQUAL'')  suffered from certain deficiencies noticed   by the authors from the outset: gravitational waves appeared to propagate with velocity greater than that of light;  gravitational lensing and cluster dynamics  could not be accounted for.  Moreover, the different conformal aspects in JBD theory,  ``Jordan frame'' and ``Einstein frame'', entered the analysis in a rather unclear way,  typical for JBD-theory  at the time.\footnote{Still in later presentations Bekenstein conceived the Jordan frame as   ``the metric measured by rods and clocks, hence the physical metric'', while Einstein frame played the role of a ``primitive metric'' which governed the Einstein-Hilbert action ``in order not ot break violently with GR \ldots'' \cite[p. 5f.]{Bekenstein:2004}. }

In the meantime it has become clear that such different, conformally related, ``frames'' are  better analyzed in terms of integrable Weyl geometry. There  they reappear as different scale gauges of the (conformal) class of pseudo-Riemannian metrics.\footnote{\cite{Quiros_ea:2013} or  \cite[sec. 3]{Scholz:2014paving}. }
But, alas,  the Weyl geometric approach to gravity is not yet well known in mainstream gravity theory. Therefore this paper starts with short introductions to (integrable) Weyl geometry (section \ref{section Weyl geometry}) and its consequences for gravity theory  (section \ref{section WgST}) in order to make it relatively self-contained.
 We then analyze how the  original AQUAL Lagrangian can be put into a scale invariant form. Scale invariance constrains  its form  strongly. In its most simple form it is given by a cubic expression in the gradient of the  scalar field.  In Einstein gauge, the  scale covariant coefficient of this term turns into  a  constant $\tilde{a}_o$ which plays  a role analogous to the MOND constant $a_o$, but is not identical with it  (section \ref{section MOND approx}). 

The conceptual  clarification achieved by this move is striking: In the  weak field, static, low velocity approximation the metrical representation of the Newton potential  is kept intact for the Riemannian component of the Weyl metric, while  the Weylian scale connection  induces an additional acceleration for the dynamics of test bodies. It has the scale invariant form of the scalar field in Riemann gauge as its  potential.  The additional acceleration is part of an extended metrical theory of gravity; it needs no other structural element (section \ref{subsection geodesics}).  Specifying these general considerations to the case of a scalar field with the cubic  Lagrangian introduced in section \ref{subsection cubic}  leads, in good approximation, to a MOND-like modified Poisson equation very much like in  RAQUAL. But here it governs only the (``anomalous'') additional acceleration induced by the Weylian scale connection, while the Riemannian component remains governed by the ordinary Possion equation (which will acquire an additional source term,  as we shall see in a moment). Conditions for the applicability of this (MOND-) approximation are estimated. In the  MOND and deep MOND regimes   the condition is satisfied for star neighbourhoods; on larger scales it even may promise a better understanding of cluster dynamics (section \ref{subsection approx conditions}).

A new feature arises from the evaluation of the energy-momentum tensor of the scalar field in the Weyl geometric framwork.  The most important contributions to the energy tensor derive from boundary terms in varying  the modified Hilbert action. Here they give rise to 
an energy density of the scalar field, which cannot be neglected for the dynamics of the systems under study  (section \ref{subsection scalar field energy}).  They add a scalar field contribution to the right hand side of the  Newtonian Poisson equation and  lead to a second addition to the Newton acceleration, proportional to the MOND acceleration of the scale connection. The effect of both additions is to be equated with the empirically determined  acceleration in the deep MOND regime (section \ref{subsection a-add}). This requires the constant  $\tilde{a}_o$ to be $\frac{1}{16} a_o$. Then the weak field, static, low velocity limit of the Weyl geometric gravity theory acquires a MONDian phenomenology. 

The Weyl geometric MOND model has  (well-determined, not freely selectable) transition functions $\mu_w(x)$ and $\nu_w(y)$ which describe the transformation from Newton acceleration to the total modified acceleration, although {\em only in the upper transition regime} to the deep MOND domain (cf. appendix \ref{appendix 3}). To my knowledge, the resulting transition functions have not yet been considered in the literature;  here they are compared with some transition functions which are in use for modelling galaxies or galaxy clusters in the astronomical literature ($\mu_1(x), \, \mu_2(x)$ and the corresponding $\nu$-functions). This comparison shows that the so-called ``phantom'' energy density  is higher in the Weyl geometric model (section \ref{subsection transition fct}).

A short discussion of the outcome of our analysis follows (section \ref{section discussion}).

\section{\small  A Weyl geometric approach to  gravity  \label{section Weyl geometry}}
\subsection{\small Some basics of Weyl geometry}
We use Weyl geometry as our geometric framework.\footnote{For more details see, among many others, \cite{Adler/Bazin/Schiffer,Blagojevic:Gravitation,%
Quiros:2013,Quiros:2014,Scholz:2011Annalen} from the point of view of  physics, for a  differential geometric perspective \cite{Folland:1970,Higa:1993,Gilkey_ea}.   \label{fn literature WG}}
It combines a {\em conformal  structure}, given by an equivalence class $\mathfrak{c}=[g]$ of pseudo-Riemannian metrics $g: ds^2 = g_{\mu \nu}dx^{\mu}dx^{\nu}$ (in local coordinates) and a {\em uniquely determined affine connection} $\Gamma$ (in local coordinates  $\Gamma_{\mu \nu}^{\nu}$) with {\em covariant derivative} $\nabla$.  The two constitutive elements of the structure $\mathfrak{c}$ and $\nabla$ (respectively $\Gamma$) satisfy the following {\em compatibility} condition: Any choice of  $g$ in  $\mathfrak{c}$ specifies  a  real valued differential 1-form $\varphi$ which depends on $g$, in coordinates $\varphi = \varphi_{\mu}dx^{\mu}$,  such that the covariant derivative of  $g$ is   $\nabla_{\hspace{-0.2em}\lambda}\,g_{\mu \nu} = - 2 \varphi_{\lambda} \, g_{\mu \nu}$, i.e. 
\beq \nabla g + 2 \varphi \otimes g  = 0\, .\label{eq compatibility} 
\eeq 
In the mathematical literature  a pair of data $(\mathfrak{c}, \nabla)$ satisfying (\ref{eq compatibility}) is called a {\em Weyl structure}.\footnote{ \cite{Higa:1993,Calderbank:2000,Ornea:2001,Gilkey_ea}.}

A change of the conformal representative
\beq g \mapsto \tilde{g}= \Omega ^2 \, g = e^{2 \omega} \, g\; , \qquad \omega = \ln \Omega , \label{eq rescaling}
\eeq 
 with diff'ble functions $\Omega$ or $\omega$, is accompanied by a change of the 1-form 
\beq \varphi \mapsto \tilde{\varphi} = \varphi-   d \ln \Omega =  \varphi -  d \omega  \; . \label{eq gauge transformation}  \eeq
This is the local description of a {\em gauge transformation} for the  connection $\varphi$ in the trivial line bundle  over spacetime of the scaling group $(\R^{+},\cdot)$.   

The change of the conformal representative $g$ has a natural physical interpretation as a point dependent {\em change of measurement units}, of scale (or ``length'') {\em gauge} as Weyl called it.\footnote{Compare with Brans/Dicke's view, most clearly expressed  in \cite[p. 2163]{Dicke:1962}.} With  basic physical units expressed in terms of time as the only elementary quantity and natural constants, like in   the new SI regulations,  the scale change of length/time units induces a coherent rescaling of the most important basic SI units.\footnote{\cite{SI:2011}, (www.bipm.org/en/si/new$_{-}$si/) }    Weyl introduced (\ref{eq gauge transformation})  as a gauge transformation of the scale connection long before the general theory of connections in principal fibre bundles was  developed, or the SI headed towards universal natural units of measurements  \cite{Weyl:GuE}. In his view the primary data of the generalized geometrical structure were given by  pairs $(g, \varphi)$ under the equivalence  ((\ref{eq rescaling}), ( \ref{eq gauge transformation})). Accordingly we call  the  equivalence class 
\beq  [(g, \varphi)] \qquad \mbox{ a {\em Weyl(ian) metric}}\,.  \label{eq Weyl metric}  \eeq
 Any specific choice of  $(g, \varphi)$ is a (scale) {\em gauge} of the Weylian metric, $g$ its {\em Riemannian component} and $\varphi$ the corresponding {\em scale connection}. 

Weyl geometry is closely related to  conformal geometry; its main difference is the {\em unique} determination of an {\em invariant affine connection} (and with it a covariant derivative). For any choice $(g,\varphi)$, the invariant affine connection may be expressed in terms of the (scale dependent)  Levi-Civita connection ${}_g\Gamma^\mu _{\nu \lambda }$ of the Riemannian component $g$ and an additional term ${}_{\varphi}\Gamma^\mu _{\nu \lambda }$ depending on the scale connection: 
\beq \Gamma^{\mu }_{\nu \lambda } =   {}_g\Gamma^\mu _{\nu \lambda } + {}_{\varphi}\Gamma^\mu _{\nu \lambda }, \qquad    {}_{\varphi}\Gamma^\mu _{\nu \lambda }=    \delta ^{\mu }_{\nu } \varphi _{\lambda } +
\delta ^{\mu }_{\lambda } \varphi _{\nu } - g_{\nu \lambda } \varphi^{\mu }.\label{Levi-Civita}  
\eeq 
The Riemann and Ricci  tensors  $Riem, \,  Ricc$ of the affine connection are {\em invariant} under scale change  although it is possible, and often  important, to express them   in terms of the scale dependent quantities $g$ and $\varphi$  in the form  $Riem=_g\hspace{-0.2em}Riem+ _{\varphi}\hspace{-0.2em}Riem  $, with  $_g\hspace{-0.1em}Riem$ the Riemannian curvature derived from the Levi-Civita connnection of $g$ and  $_{\varphi}\hspace{-0.1em}Riem$ the correction term derived from the scale connection $\varphi$; similarly $Ricc=_g\hspace{-0.2em}Ricc+ _{\varphi}\hspace{-0.2em}Ricc  $.\footnote{For explicit formulas see the literature given in fn. \ref{fn literature WG}}

The Weyl geometric scalar curvature $R=g^{\mu \nu}R_{\mu \nu}$ is not  scale invariant but scales with $g^{\mu \nu}$ (weight $-2$, cf. below). It is composed  from  the scalar curvature of the Riemannian component $_gR$  and a  term collecting the influence of the scale connection $_{\varphi}R$   
\beqarr
  R &=& _g\hspace{-0.3em}R+ _{\varphi}\hspace{-0.3em}R \label{eq R}\\
_{\varphi}\hspace{-0.1em}R  &=& - (n-1)(n-2) \varphi_{\lambda } \varphi^{\lambda }  - 2(n-1) _g \hspace{-0.2em}\nabla _{\lambda }\varphi^{\lambda } \nonumber \\ 
&=& - 6 \varphi_{\lambda } \varphi^{\lambda }  - 6 _g \hspace{-0.2em}\nabla _{\lambda }\varphi^{\lambda } \;  \qquad \mbox{in dimension $n=4$.}  \nonumber
 \label{eq scalar curvature}
\eeqarr

Of course, the scale connection has a curvature  $f$ of its own. Because  the  commutative scale group  it is simply the exterior derivative
\beq   f = d \varphi \qquad \mbox{\em (scale curvature).} \label{eq scale curvature}
  \eeq
	If it  vanishes, $d\varphi=0$, there is a scale choice of the Weylian metric, $(\tilde{g},0)$, in which  the scale connection vanishes  ({\em integrable Weyl geometry}).  Then the  Weyl metric {\em looks}  Riemannian in this gauge; but it would be a mistake to identify it  with the Riemannian metric $g$ because   the underlying scale covariance group is {\em not  reduced} to the identity. Even in the  case of an {\em integrable Weyl geometry} the group of {\em geometrical automorphisms} contains the {\em conformal} transformations. It is important to keep this (simple) observation in mind for the study of  scalar tensor theory of gravity in the Weyl geometric framework.\footnote{In his reflections on the quantization of gravity  't Hooft  considers ``local conformal symmetry''  as an exact symmetry, although explicitly avoiding  to make use of the Weyl geometric framework \cite[fn. 2]{tHooft:2014}. Perhaps it would be helpful to give up this methodological restriction.}

Some geometrical and many physical quantities are given by fields $X$ which transform under rescaling. Mathematically speaking, such fields live (i.e. have values) in bundles over spacetime with non-trivial representation of the scale group. A field $X$ transforming by $ \tilde{X}= \Omega ^k X = e^{k \,\omega}X$  under  
 (\ref{eq rescaling}) is known as {\em scale covariant} field of Weyl {\em weight} $k$. For geometrical reasons we work with length/time weights, inverse to energy weights preferred in high energy physics  by obvious reasons. 
The {\em scale covariant derivative} $D$ of such a field $X$ responds to the non-trivial weight; it is given by
\beq DX := \nabla X + w(X)\varphi  \otimes X \; . \label{scale covariant derivative} \eeq
We now see that the compatibility (\ref{eq compatibility}) means
$ D g = 0$,  
i.e. the {\em scale covariant derivative of the metric vanishes} --  a Weyl geometric analogue of the metricity condition for the Levi-Civita connection in Riemannian geometry. 

In addition to the notations $\nabla$ for {\em scale  invariant} covariant differentiation and $D$ for {\em scale covariant} differentiation of fields we shall use the notation  $_g\hspace{-0.1em}\nabla$ for the scale dependent differentiation with regard to the Levi-Civita connection of the Riemannian component $g$ of a Weyl metric given in gauge $(g,\varphi)$.

Weyl geometry connects to physics via different routes. Leaving aside Weyl's own idea of a  geometrically unified theory of electromagnetism and gravity, two different research programs developed in the second half of the 20th century. The first one  in the theory of gravity (with links to elementary particle physics and cosmology)  characterized by a gravitational scalar field non-minimally coupled to the scalar curvature, similar to Jordan-Brans-Dicke theory (going back to M. Omote and P.A.M. Dirac in the early 1970); the second one arising from a Weyl geometric re-reading of Bohmian quantum mechanics with a scale covariant scalar field in the role of a generalized quantum potential (opened by E. Santamato in the 1980s).\footnote{For the quantum potential approach see, among many, \cite{Santamato:WeylSpace,DeMartini/Santamato:Dirac_equation,DeMartini/Santamato:quantum_nonlocality,Shojai/Shojai:2002,Carroll:quantum_potential}.} In recent years the gravitational scalar field approach has been taken up  in the simplified form of {\em integrable} Weyl geometry. Our investigation is part of this  research tradition.

\subsection{\small  Weyl geometry as a framework for gravity  \label{subsection framework}}
 Lagrangians of field theories in the Weyl geometric framework have to be invariant under scale transformation (conformal invariance). It is advisable to express them in terms of the scale co- or invariant expressions outlined above. 
Weyl himself worked with quadratic expressions in the curvature to get scale 
invariant Lagrangians. A similar approach is still used in conformal theories of gravity.\footnote{ \cite{Mannheim:2006}}
But roughly a decade after the advent of Brans-Dicke theory several authors, beginning with  M. Omote and P.A.M. Dirac,  formulated a Weyl geometric version of a scalar field  $\phi$ of weight $w(\phi)=-1$ non-minimally coupled to Weylian scalar curvature $R$, with a {\em Hilbert-Weyl} term $L_{HW}= |\phi|^2 R $.\footnote{\cite{Omote:1971,Dirac:1973,Omote:1974,Utiyama:1975I,Utiyama:1975II,Hayashi/Kugo:Weyl_field}.}  
Originally the Weylian scale connection was treated as a  dynamical field  with a Yang-Mills like Lagrange term for $\varphi$.\footnote{Dirac continued to interpret $\varphi$ as electromagnetic potential, while the Japanese physicists hoped for a new insight into nuclear fields.}

It was  soon realized that such a field would have a boson  close to the Planck scale.  Some authors  speculated that the scalar field might arise as an order parameter  of a boson condensate.\footnote{\cite{Hayashi/Kugo:Weyl_field,Smolin:1979,HungCheng:1988,Hehl_ea:Local_scale_1998}.} In such a case,  the low energy effective Lagrangian does not attribute an independent dynamical role to the scale connection because the scale curvature vanishes for low energies.\footnote{Curvature effects can be seen only at lengths/energies close to the Planck scale.}
 The only additional dynamical effect of the field theoretic extension is due to the scalar field. A {\em geometrical} role of the scale connection remains even in this  case of an integrable Weyl geometry. All this  is consistent with the outcome  of  Ehlers/Pirani/Schild's  analysis on the foundational role of Weyl geometry,  and the succeeding investigations of Audretsch/G\"ahler/Straumann.\footnote{\cite{EPS} show that the causal structure and  a compatible non-chronometric inertial structure (mathematically  a  conformal and a compatible path structure) uniquely specify a Weylian metric. \cite{AGS} have shown that, in the WKB approximation, the streamlines of a  Klein-Gordon field approximate the geodesics of the Weyl metric if and only if the scale curvature vanishes.}

We  arrive at   a scalar tensor theory of gravity (and other fields) with a Lagrangian of the general form
\beqarr L &=& \alpha \phi^2 R +  \ldots \label{Lagrangian general form}\\
   \mathfrak{L} &=& L \sqrt{|g|}\, , \qquad |g|=   |det\, g|\, , \nonumber
\eeqarr 
where the dots  indicate scalar field, matter and interaction terms. Obviously (\ref{Lagrangian general form})  is very  close to Jordan-Brans-Dicke theory (JBD). 
The crucial difference is  that in our case 
 the scalar curvature $R$ and all  dynamical terms are consistently expressed in Weyl geometric scale covariant form and the  Lagrangian remains scale (conformally) invariant for any $\alpha$, not only for $\alpha = \frac{1}{6}$.  Scale covariance has not to be broken by hand. There are no ``two'' (or even more) ``metrics'' involved. The  notorious question of ``physicality'' of frames in JBD theory is brought into a different (and clarifying) light.\footnote{\cite{Quiros_ea:2013,Quiros:2014,Poulis/Salim:2011,Romero_ea:Weyl_frames,Almeida/Pucheu:2014,Scholz:2014paving}.}
In short, the  Weyl geometric framework  brings in  more clarity of concepts and simplifies calculations.  

\subsection{\small Scale invariant observables and two distinguished gauges \label{subsection scale invariant observables}}
It is clear how to extract  {\em scale invariant observable magnitudes}  $\check{X}$ from a scale covariant field $X$ of weight $w(X)=k$. One only has to form the  proportion with regard to the appropriate power of the scalar field's norm 
\beq \check{X}:= X / \, |\phi|^{-k}  = X |\phi|^k\, ;  \eeq
then clearly $w(\check{X})= 0$.

 Scale invariant magnitudes $\check{X}$  are directly indicated, up to a globally constant factor in {\em scalar field gauge}, i.e., the gauge in which  
\beq
|\phi| \doteq const =: \phi_o\,, \label{phi-o}
\eeq
where  $  \doteq$ indicates an {\em equality which holds in a specified  gauge only} (here scalar field gauge).
In  \cite{Utiyama:1975I}   $\phi$ is therefore called a ``measuring field''. In our context $\phi$ will be strictly positive real valued; thus we can omit the norm signs in the expressions above.  By  consistency considerations  with Einstein gravity we have to postulate that in scalar field gauge 
\beq \alpha \phi^2 \doteq  \alpha \phi_o^2 = (16 \pi G)^{-1}\, , \label{eq Einstein gauge}
\eeq   Scalar field gauge with (\ref{eq Einstein gauge}) will be called and denoted  by
\beq (\hat{g}, \hat{\varphi}) \qquad \mbox{\em Einstein (--  scalar field) gauge.}
\eeq 
Once the context is clear, the hats may be (and will be)  omitted.

In integrable Weyl geometry there is another distinguished gauge of the form    $(\tilde{g},0)$ in which the scale connection vanishes. By obvious reasons it is called
\beq  (\tilde{g},0) \qquad  \mbox{ \em Riemann gauge}
\eeq 
(``Jordan frame'' in JBD theory). Writing the scalar field in Riemann gauge $\tilde{\phi}$ in exponential form,
$\tilde{\phi} = e^{\tilde{\omega}}$,  turns its exponent
\beq
\tilde{\omega}:= \ln{\tilde{\phi}} \label{omega}
\eeq
into  a {\em scale invariant} expression for the scalar field. (Further below, we shall omit the tilde sign, if the context makes clear that the scale invariant exponent is meant.) The scale connection $\varphi = \hat{\varphi}$ in scalar field gauge  is then 
\beq \hat{\varphi} = - d  \tilde{\omega} \, , \label{eq check varphi}
\eeq 
because $\Omega=\tilde{\phi}$ is the rescaling function from Riemann to scalar field gauge.

Riemann gauge and scalar field/Einstein gauge    are the most important  gauges in Weyl geometric scalar field theory. In the first one, the affine connection is identical to the Levi-Civita connection of the Riemannian component $\tilde{g}$.\footnote{Some authors in the JBD approach consider this as the criterion for the ``physical'' gauge \cite{Bekenstein:2004}.}
 In the second one, the coefficient of scalar curvature is consistent with Einstein gravity and the scale invariant observables are directly indicated by the field quantities without further calculation (up to a global constant). We may expect, or postulate, that clock measurements are indicated by quantities in this gauge.\footnote{For a possible physical reason,   mediated by a link to the Higgs field,  see \cite{Scholz:Weyl_gauge}.} 
Thus both gauges have their mathematical {\em and physical} values and vices; both indicate some physically important feature most directly, while others have to be extracted by additional calculations. Both are equivalent mathematically. 

\subsection{\small Inertio-gravitational, conformal, and chronometric structures \label{subsection geodesics}}
{\em Scale invariant geodesics} are the autoparallels of the scale invariant derivative, i.e. paths $\gamma(t)$ satisfying
\beq \nabla_{\dot{\gamma}}(\dot{\gamma}) = 0\, \qquad \longleftrightarrow  \quad \ddot{\gamma}^{\lambda} + \Gamma_{\mu \nu }^{\lambda}\dot{\gamma}^{\mu}\dot{\gamma}^{\nu} = 0\; .  \label{eq invariant geodesic}
\eeq 
The corresponding {\em scale covariant geodesics} arise from (\ref{eq invariant geodesic}) by reparametrizing these paths to unit length in any gauge. Their vector fields $u(t)=\dot{\gamma}(t)$,  defined along every path, are of weight $w(u)=-1$;  then we have $g(u,u)=\pm 1$ independent of the scale gauge. They are given by 
\beq D_{u} u =\nabla_{u} u - \varphi(u)u  = 0 \qquad \longleftrightarrow  \quad \dot{u}^{\lambda} + \Gamma_{\mu \nu }^{\lambda}u^{\mu}u^{\nu} - \varphi_{\mu}u^{\mu}u^{\lambda} = 0\; . \label{eq covariant geodesic}
 \eeq 
The autoparallels of (\ref{eq covariant geodesic}) differ from Weyl's scale invariant geodesics (\ref{eq invariant geodesic}) by parametrization only and constitute a class of {\em covariantly parametrized geodesics}.\footnote{More generally, a path $\gamma$ in a Weylian spacetime manifold $M$ is called {\em scale covariantly  parametrized} of weight $-1$, if to any scale choice $(g,\varphi,\phi)$  a  parametrization $\gamma:  \R \longrightarrow M$ is given, which changes under rescaling of the metric in such a way that $g(\dot{\gamma(\tau)},\dot{\gamma(\tau)})$ is independent of the gauge.}
They are the autoparallels of a projectively  related class $[\tilde\Gamma(\varphi)]$ of affine connections $\tilde\Gamma(\varphi)$ depending on the gauge $(g, \varphi)$: 
\beq  \tilde{\Gamma}(\varphi)_{\mu \nu }^{\lambda} =  \Gamma_{\mu \nu }^{\lambda} -  \underline{\frac{1}{2}(\delta^{\mu}_{\nu} \varphi_{\kappa} + \delta^{\mu}_{\kappa} \varphi_{\nu})} \label{eq projective class}
\eeq 
Here the additional term arsing from scale covariant derivation of weight -1 has been underlined. 
The class $[\tilde{\Gamma}]$ characterizes a {\em projective} path structure $[\gamma]$ with paths given by (\ref{eq covariant geodesic}).\footnote{That  (\ref{eq invariant geodesic}) and (\ref{eq covariant geodesic}) characterize the same path structure can be verified by the criterion of {\em projective equivalence} for two connections $\Gamma,  \tilde{\Gamma}$, which is $(\tilde\Gamma -\Gamma)^{\mu}_{\nu \kappa}X^{\mu}X^{\kappa} \sim X^{\mu}$ for any vector field $X$.}  

According to the analysis of Ehlers/Pirani/Schild the projective and the conformal structure $\mathfrak{c}$ specify the affine connection and its covariant derivative  $\nabla$ uniquely. As also the Weyl structure specifies the projective structure we  have three equivalent characterizations of a Weyl geometry:
\beq ( \mathfrak{c}, [\tilde{\Gamma}])   \quad \longleftrightarrow \quad (\mathfrak{c}, \nabla) \quad \longleftrightarrow \quad  [(g,\varphi)]\, ,
\eeq 
with $[(g,\varphi )]$ a Weylian metric in the sense of (\ref{eq Weyl metric}).
Each of them defines an {\em inertio-gravitational} structure in the sense of Weyl while the chronometry is still undetermined up to a point dependent scale factor.  

As shown in  section \ref{subsection scale invariant observables}, a scale covariant scalar field $\phi$  as in section \ref{subsection framework} specifies a  {\em chronometry}. A Weylian metric plus  a scalar field  $[(g,\varphi,\phi)]$ thus determine a full-fledged {\em spacetime structure} in the sense of \cite{Stachel:2003space-time}. Remember that in the case of an integrable Weyl structure $\varphi$ and $\phi$ are not dynamically independent but determine each other mutually. 
Any Weyl geometric  scalar field  theory contains point dependent rescaling as a subgroup of its automorphisms. The choice of Einstein - scalar field gauge allows to specify the chronometric structure in an adapted way but does not reduce the group of automorphisms.

 \subsection{\small Additional acceleration induced by the scale connection \label{subsection acceleration scale connection}}
 Free fall of test particles  in Weyl geometric gravity follows scale covariant geodesics $\gamma(\tau)$ of weight $w(\dot{\gamma}) = -1$.
 Slow (non-relativistic) motions are described by a differential equation  formally identical to the one in Einstein gravity, but with scale covariant derivatives of the Weyl geometric affine connection rather than that of the (Riemannian) Levi-Civita one.

Coordinate acceleration $a$ with regard to proper time $t$ for a low velocity motion parametrized by $x(t)$ is given (analogous to  Einstein gravity) by\footnote{\cite[pp. 213ff.]{Weinberg:Cosmology_1972} or, for Weyl geometry, \cite[eq.(60)]{Scholz:model_building_2005}.} 
\beq a^{j}= \frac{d^2x^{j}}{dt^2} \approx - \Gamma^{j}_{oo}\, . \label{acc 1}
\eeq 
Because of (\ref{Levi-Civita}) the total acceleration decomposes into 
\beq a^{j} = - _g\hspace{-0.1em}\Gamma^{j}_{oo} - _\varphi\hspace{-0.2em}\Gamma^{j}_{ \nu \lambda} =a^{j}_R + a^{j}_{\varphi} \, , \label{acceleration}
\eeq 
where $a^{j}_R=- _g\hspace{-0.1em}\Gamma^{j}_{oo} $ is the Riemannian  component of the acceleration known from Einstein gravity, and  $a^{j}_{\varphi}= - _\varphi\hspace{-0.15em}\Gamma^{j}_{oo}$   an {\em additional acceleration}   due to the   Weylian scale connection.

For a (diagonalized) weak field approximation in Einstein gauge, 
 \beq 
 g_{\mu \nu} \doteq \eta_{\mu \nu} + h_{\mu \nu}, \qquad |h_{\mu \nu}| \ll 1\, , \label{weak field}
 \eeq 
 with $\eta = \epsilon_{sig} \, \mbox{diag}(-1,+1,+1,+1)$, the Riemann-Einstein component is standard:
 \beq
a^{j}_R=- _g\hspace{-0.1em}\Gamma^{j}_{oo} \; \dot{\approx} \;  \frac{1}{2} \eta ^{j j}\partial_{j} h_{oo} \, , \label{Newton limit}
 \eeq
neglecting 2-nd order terms in $h$.
   In the light of (\ref{Levi-Civita}) and (\ref{omega potential}) the  additional Weylian component becomes  
 \beq
 a^{j}_{\varphi} \doteq g_{oo}\varphi^{j}   \doteq g_{oo}g^{jj}\partial_{j} \tilde{\omega} \;  \dot{\approx} \;  - \partial_{j} \tilde{\omega} \;  \dot{\approx} \; \varphi_{j} \,,
 \eeq
 This shows that in the static weak field, low velocity  case and in Einstein gauge the Weylian {\em scale connection} represents an {\em additional acceleration}. 

Because of (\ref{eq check varphi})  the  {\em invariant} form of the {\em scalar field}  $\tilde{\omega}$ can be identified with the {\em potential of the additional acceleration} (weak field approximation, Einstein gauge), analogous to Einstein's identification of the Newton potential with a metrical perturbation,   $\Phi_N := -\frac{1}{2}\epsilon_{sig} h_{oo}$:
\beqarr 
a_R &\dot{\approx}& - \nabla \Phi_N  =  -\frac{1}{2}\epsilon_{sig}  \nabla h_{oo} \label{a_R}\\
a_{\varphi} &\dot{\approx}& - \nabla \tilde{\omega} \label{a_w}
\eeqarr

\section{\small Weyl geometric scalar tensor theory (W-ST) \label{section WgST}}

\subsection{\small \ldots with a cubic scalar field Lagrangian \label{subsection cubic}}

 Our Lagrangian density $\mathcal{L}=L\sqrt{|g|}$ contains a Hilbert-Weyl term $ L_{HW}$, a dynamical term  $L_{\phi}$ and a potential term $L_{V4}$  for the scalar field and a   matter term $L_m$, all of them of weight $-4$:
  \[
  L = L_{HW}+L_{V4} + L_{\phi}+L_m \; 
  \]
We assume a classical matter term with $w(L_m)=-4$ comparable to the matter terms of the standard model fields, for which  test particles follow the {\em Weyl geometric path structure}.
The postulate  is strongly supported by the analysis of the stream lines of a Klein-Gordon field (in WKB approximation)  \cite{AGS},  if one assumes  a 
structure-conserving transition from the quantum world to classical particle  motion after decoherence. It can be understood as a compatibility criterion of the matter Lagrangian  with the EPS  axioms for a generalized theory of gravity (Ehlers/Pirani/Schild).\footnote{ This assumption  deserves further investigation. It can be stated as an action principle for point particles with the scale invariant action: $S_{pp}=\int \phi_{comp}\sqrt{g(\dot{\gamma}\dot{\gamma}) }\,d\tau$  (with  $\gamma$ timelike curves parametrized by $\tau$, $\phi_{comp}$ the ``compensating field'' like in appendix \ref{appendix 1}); but the question of consistency or derivability  would still persist. In  \cite{Almeida/Pucheu_ea:2014} it is derived for a weak extension of Einstein gravity, rewritten scale covariantly using Weyl geometry  (by means of the contracted Bianchi identity applied to the energy-momentum of  dust-like matter, like in ordinary Einstein gravity). This approach might be generalizable. The condition of {\em EPS compatibility} is analyzed in great generality in \cite{DiMauro_ea:Further_theories_I}.}  

For covering  both  signature choices  for $g$,  preferentially used in gravity theory or in elementary particle physics, we introduce
  \beq \epsilon_{sig}= \left\{ {+1 \quad \mbox{if sign}(g) =(3,1) \sim (-+++)} \atop {-1 \quad \mbox{if  sign}(g) =(1,3) \sim (+---)} \right.
  \eeq 
and a modified Hilbert term typical for  scalar-tensor theories of gravity, adapted to the Weyl geometric framework:
 \beqarr L_{HW} &=& \frac{\epsilon_{sig}}{2} (\xi \phi)^2 R \qquad \qquad \mbox{Hilbert-Weyl term},  \label{Lagrangian}\\
 L_{V4} &=& - \frac{\lambda}{4}\, \phi^4 \qquad \qquad \qquad \mbox{quartic potential term of} \; \phi,
\eeqarr
 with constants $\xi, \, \lambda$  to be interpreted later.   $R$ is the Weyl geometric scalar curvature, scale covariant of weight $w(R)=-2$. The coefficient $\xi$ has to be fixed such that in scalar field/Einstein gauge $ (\xi \phi)^{-2} \doteq 8 \pi G$. So far all {\em Weyl geometric scalar tensor theories} of gravity (W-ST) coincide.  

Usually the dynamical term $L_{\phi}$ of the scalar field is quadratic in its  (scale covariant)  gradient, i.e. proportional to
$ (D\phi)^2=D_{\nu}\phi D^{\nu}\phi\, . $ In order to adapt to our form of the Hilbert term we write it in the form
\beq L_{\phi\,2} = \epsilon_{sig} \frac{\alpha}{2}\xi^2 |D\phi|^2 =  \epsilon_{sig} \frac{\alpha}{2}\xi^2 D_{\nu}\phi D^{\nu} \phi \; . \label{L_phi2 a}
\eeq 

But here we want to reconsider the alternative of an {\em aquadratic} Lagrangian proposed by  Bekenstein/Milgrom for reproducing the non-linear Poisson equation of the MOND phenomenology in the static weak field limit,\footnote{\cite{Bekenstein/Milgrom:1984}} 
\beq
(8 \pi G)^{-1} c^{-2} \, f(c^2 \, (\nabla\phi)^2)\, , \label{eq AQUAL}
\eeq
 where $f$ is a non-linear function and the constant $c$ has ``dimensions of length introduced for dimensional consistency'' \cite[p. 6]{Bekenstein:2004}. 
Bekenstein's $f$ could be chosen among a large class of functions (it is not ``not known {\em apriori''}) and is functionally related to the MOND specific transition function $\mu(x)$ from the Newton regime to the  deep MOND domain. That implies the asymptotic condition 
\beq
f(y)\rightarrow y^{\frac{3}{2}} \quad \mbox{ (up to a constant factor) for} \qquad y \rightarrow 1 \,.  \label{asymptotic condition}
\eeq
Assimilating (\ref{eq AQUAL}) to our context, $f$ will be strongly 
 constrained by  the total weight condition $w(L_{\phi})=-4$ and the asymptotic condition (\ref{asymptotic condition}). The simplest non-quadratic form is  $f(y) = y^{\frac{3}{2}}$ itself (for $y\geq 0$), with a reduction of  the exponent of the factor $c^{-2}$ in front of $f$ in Bekenstein's Lagrangian to $-1$.

For achieving scale invariance of $\mathcal{L}_{\phi}$ we set
\beq
 L_{\phi\, 3} = \frac{2}{3}\, \xi^2\eta \,\phi^{-2}\, \left\| D\phi \right\|^3\; \\
 \label{L_phi 3} 
 \eeq 
and add it to $ L_{\phi\, 2}$
for the  kinetic term of   $\phi$.  $D\phi$ denotes the {\em scale covariant gradient} of $\phi$ with components $D^{\nu}\phi$;  the ``norm''  $\left\|X \right\|$ of a 4-vector $X=(X^{\nu})$ is  to be read as
\beq \left\| X \right\|=Re\,(\epsilon_{sig}X^{\nu}X_{\nu})^{\frac{1}{2}}\, .  \label{|| ||}
\eeq 
For spacelike vectors it is the usual norm, for timelike vectors it is zero.\footnote{As an alternative convention on might consider $\left\| X \right\| = |X^{\nu}X_{\nu}|^{\frac{1}{2}}$. Consequences of this alternative convention, e.g.  for cosmological solutions or propagation of perturbations, are still to be explored.}
The coefficient $\eta$ allows  to  adapt the model  to Bekenstein/Milgrom's  value of their constant $a_o$. 
The factor $\frac{2}{3}$ is for convenience. The scale weight of $\left\|D\phi \right\|$ is $-2$, thus $w(L_{\phi})=2- 3\cdot 2 = -4$. The condition of scale invariance for $\mathfrak{L}_{\phi}$ constrains Bekenstein/Milgrom's  $f$ considerably. 

Adapted to (\ref{|| ||}) there is now also the  possibility  to consider 
\beq L_{\phi\,2} = \epsilon_{sig} \frac{\alpha}{2}\xi^2 ||D\phi||^2    \label{L_phi2 b}
\eeq 
 as an alternative for the quadratic term, while in any case 
\beq    L_{\phi} =  L_{\phi\,2} +  L_{\phi\,3} \, . \label{L_phi}
\eeq 
Although $ L_{\phi\,2}$ is scale covariant of weight $-4$ for any choice of $\alpha$, the specific choice $\alpha=6$ leads to the effect that in vacuum the scalar field equation derived from the quadratic term $ L_{\phi\,2}$ alone reduces to the trace of the Einstein equation. This property will allow to simplify the total scalar field equation of our $L_{\phi}$ considerably (subsection \ref{subsection dynamics}).
 
The gradient of the scalar field in terms of its invariant form $\tilde{\omega}$ (\ref{omega}) is $D^{\nu}\phi = \phi\, \partial^{\nu}{\tilde{\omega}}$ (appendix \ref{appendix 1}, eq. (\ref{D nu phi})). Thus the   scalar field  Lagrangian can also be written with
 \beq L_{\phi\, 3} = \frac{2}{3}\xi^2 \eta \, \phi \, \left\|\nabla \tilde{\omega}\right\| ^3 \, , \qquad                     \label{cubic Lagrangian}
 \eeq 
with $\nabla \tilde{\omega}$ the gradient of $\tilde{\omega}$. 
In Einstein gauge (\ref{eq atilde}),  with constant value $\phi_o$ of the scalar field, we introduce the new constant $ \eta^{-1}\phi_o = \tilde{a}_o$.\footnote{Then $\xi^2 \eta \, \phi = (\xi \phi)^{2} (\eta^{-1}\phi)^{-1}\doteq (8 \pi G)^{-1} \tilde{a}_o^{-1}$ (in Einstein gauge).}   
 Below it will turn out that  this will be  realized with $\tilde{a}_o= \frac{a_o}{16}$.  
$L_{\phi\,3}$ is    {\em cubic} in the gradient of the scale invariant scalar field rather than quadratic (and of the correct weight because of $w(\left\| \nabla {\tilde{\omega}} \right\|)=-1$). In the following we shall omit the tilde and simply write $\omega$ for the latter.
 
\subsection{\small Compatibility conditions \label{subsection compatibility}}
Our Lagrangian is consistent with Einstein gravity if in  scalar field gauge
\beq \xi \phi_o  \doteq (8\pi G)^{-\frac{1}{2}} = E_{pl}\leftrightarrow L_{pl}^{-1}\, , \label{Einstein gauge}
\eeq
 where  $E_{pl}$, $L_{pl}$ denote the {\em reduced} Planck energy and  Planck length,  respectively. They are normed such that 
 \beq E_{pl}\, L_{pl}^{-1} = (8\pi G)^{-1} \, .  \label{reduced Planck quantities}
 \eeq   Obvious factors  $c$  and $\hbar$ are omitted.
  Einstein gravity arises if in scalar field gauge  $\varphi \rightarrow 0$.

Let us introduce the notation
\beq
   \tilde{a}_o  =  \eta^{-1}\phi_o  \,  \label{eq atilde}
\eeq
 with $\phi_o$ as in (\ref{phi-o}). The constant $\tilde{a}_o $ plays a role analogous to the MOND acceleration $a_o \approx \frac{1}{6} H$, where $H$ denotes the Hubble parameter ($H=H_o \leftrightarrow H_1$). Below we find that we have to set $\tilde{a}_o \approx \frac{a_o}{16}$ if we want to link up to Bekenstein/Milgrom's RAQUAL  with 
 the usual MOND acceleration.  Einstein gravity is (precisely)  contained in our  appoach as the special case with ${\omega} =  const $. Then Riemann gauge and Einstein gauge coincide and the scalar field is dynamically inert.\footnote{\cite[sect.3]{Scholz:Weyl_gauge}, \cite{Romero_ea:Weyl_frames}.} 
In the following we shall understand by {\em Einstein gauge}  the scalar field gauge with  (\ref{Einstein gauge}) {\em and}  (\ref{eq atilde}).

$\phi_o^{-1}$ stands between the  largest and smallest physically conceivable length units in the universe $\tilde{a}_o^{-1}$ and $L_{pl}$; or reciprocally:
\[ \tilde{a}_o \quad \stackrel{\cdot \eta}{\longmapsto} \quad  \phi_o \quad \stackrel{\cdot \xi}{\longmapsto} \quad  E_{pl} \leftrightarrow L_{pl}^{-1}
\] 
The product of our typical coefficients is  the  ratio of these extremal quantities:
\beq \eta \cdot \xi = \frac{E_{pl}}{\tilde{a}_o}  = \frac{\tilde{a}_o^{-1}}{L_{pl}}   \sim 10^{63}
\eeq 
It seems natural (although not necessary) to assume $\xi$ and $\eta$  to be at roughly comparable orders of magnitude. Then $\phi_o$ lies  close to the geometrical mean  between the extremes $\tilde{a}_o$ and $ E_{pl}$: 
\beq |\phi_o| \sim  10^{-4}\, eV  \quad \mbox{respectively} \quad  10\,  cm^{-1} \label{geometrical mean}
\eeq  

\subsection{\small Dynamical equations \label{subsection dynamics}}
 In integrable Weyl geometric structures the scale covariant variation with regard to $\delta g^{\mu \nu}$ leads to the Euler-Lagrange equation
\[ \frac{\delta \mathfrak{L}}{\delta g^{\mu \nu}} = \frac{\partial \mathfrak{L}}{\partial g^{\mu \nu}}-D_{\lambda}\frac{\partial \mathfrak{L}}{\partial(\partial D_{\lambda} g^{\mu \nu})}
\]
with $ D_{\lambda} g^{\mu \nu}= \partial_{\lambda} g^{\mu \nu} - 2 \varphi_{\lambda}  g^{\mu \nu} $ \cite[p. 526]{Frankel:Geometry}.  Because of $D_{\lambda}g^{\mu \nu} \doteq \partial_{\lambda} g^{\mu \nu}$ (in Riemann gauge) the variation is most simple  in  Riemann gauge  and close to the usual calculations. The result can be generalized to other gauges   by scale transformation.\footnote{For the variation in general, not necessarily integrable, Weyl geometric structures see \cite[pp. 98--101]{Tann:Diss}.}

The variation with regard to $\delta g^{\mu\nu}$ leads to boundary contributions  from the Hilbert-Weyl term, which vanish for a constant coefficient like in Einstein gravity:\footnote{\cite[pp. 96ff.]{Blagojevic:Gravitation},  \cite[pp. 40ff.]{Fujii/Maeda}, \cite[pp. 64ff.]{Tann:Diss}, \cite[p. 1032f.]{Drechsler/Tann}  -- the boundary terms lead to the  ``improved'' energy-momentum tensor of the scalar field in the sense  of \cite{Callan/Coleman/Jackiw}. \label{fn Blago}}
\beq \frac{1}{\sqrt{|g|}} \frac{\delta \mathcal{L}_{HW}}{\delta g^{\mu\nu}}= \frac{\epsilon_{sig}}{2}\xi^2\left( \phi^2(Ric-\frac{R}{2}g)_{\mu \nu} -D_{(\mu}D_{\nu)}\phi^2 + D^{\lambda}D_{_\lambda}\phi^2g_{\mu\nu} \right) \label{variation Hilbert-Weyl term}
\eeq 
Here $Ric$ and $R$ are the Weyl geometric Ricci tensor and scalar curvature respectively.  The last two terms on the r.h.s. result from the boundary contributions of  partial integration.  
Remember that $D_{\mu}$ denotes the scale covariant derivative of Weyl geometry, depending on the scale weight $w=w(X)$ of a field $X$ (\ref{scale covariant derivative}).

 The variation of the other terms is straight-forward. The energy-momentum tensor of matter is defined as usual:
 \beq T^{(m)}_{\mu \nu} := - \epsilon_{sig} 2 \frac{1}{ \sqrt{|g|}}\frac{\delta \mathcal{L}_{m}}{\delta g^{\mu\nu}}
 \eeq 
 The variation of $\mathcal{L}_{\phi}$ gives a peculiar energy-momentum contribution from the scalar field to the r.h.s. (see below, (\ref{ThetaI}), (\ref{ThetaII})). 

We arrive at the  {\em scale invariant Einstein equation}, 
\beq
 Ric-\frac{R}{2}g = (\xi \phi)^{-2}\, T^{(m)} + \Theta\, . \label{Einstein equation}
\eeq 
The r.h.s. consists of the energy-momentum of matter $T^{(m)}$ and  the energy tensor of the scalar field $\Theta$  (up to the constant $8\pi G$ in Einstein gauge). $\Theta$ decomposes  into a term (I) manifestly proportional to the Riemannian component of the metric $g$ and an additional  one (II),    $\Theta = \Theta^{(I)}+\Theta^{(II)}$,  such that 
\beqarr
\Theta^{(I)} &=& \phi^{-2} \left(  - D_{\lambda}D^{\lambda}\phi^2  +  \epsilon_{sig}\, \xi^{-2}(L_{V4} +  L_{\phi}) \right)g   \, , \label{ThetaI}\\
\Theta^{(II)}_{\mu \nu} &=& \phi^{-2} \left( D_{\mu}D_{\nu}\phi^2 
- 2 \epsilon_{sig} \xi^{-2} \frac{\partial L_{\phi}}{\partial g^{\mu \nu}} \right) \, .  \qquad    \label{ThetaII}
\eeqarr
 The contribution   $\epsilon_{sig}(\xi\phi)^{-2}L_{V4}\,g = - \epsilon_{sig} \frac{\lambda}{4} \xi^{-2}\phi^2 \, g$  in (\ref{ThetaI}) is a scale covariant version of the ``cosmological constant'' term $\Lambda  g$;  here
\beq
\Lambda =  \frac{\lambda}{4} \xi^{-2}\phi^2 \quad \mbox{(variable)} \; , \qquad \Lambda \doteq  \frac{\lambda}{4} \xi^{-2}\phi_o^2 \quad \mbox{(constant in Einstein gauge)}  \, . \label{cosmological constant}
\eeq

For the variation $\delta \omega$   with regard to the scale invariant form  of the scalar field $\omega$ one uses (\ref{phi omega}) (valid in any gauge)    and  finds
\beq \frac{\partial}{\partial \omega} \phi =  \frac{\partial}{\omega}  e^{\omega + \int \varphi} =  e^{\omega + \int \varphi} = \phi \, ,  \label{partial omega of phi}\, .
\eeq
On the other hand
\beq \frac{\partial}{\partial(\partial_{\nu} \omega)}\left\| \nabla \omega \right\|^3 = 3  \epsilon_{sig} \left\| \nabla \omega \right\| \partial^{\nu} \omega \,  \, \label{double partial omega}
\eeq
for $\nabla \omega$ spacelike; otherwise it  is 0. 
 
Let us introduce  the {\em scale covariant} (non-linear) {\em Milgrom operator} defined by
\beq \square_M\, \omega =  \epsilon_{sig} D_{\nu}(\left\| \nabla \omega  \right\| \partial^{\nu}\omega)  \, , \label{Milgrom operator}
\eeq 
and   $\square$,  the {\em scale covariant d'Alembert operator}  for a (scale covariant) scalar field $X$, while    $_g\hspace{-0.2em}\square$ is the covariant d' Alembert operator of the Riemannian metric in any gauge:
\beqarr \square X &=& -\epsilon_{sig} D_{\nu}D^{\nu}X  \, \label{d Alembert}\\
 _g\hspace{-0.05em}\square\, X &=&  -\epsilon_{sig} \, _g\hspace{-0.1em}\nabla_{\hspace{-0.1em}\nu}\,\partial^{\nu} X =- \frac{\epsilon_{sig} }{\sqrt{|g|}}\partial_{\nu}\left(\sqrt{|g|} \, X^{\nu}\right) \,  .\label{Beltrami-Laplace}  
\eeqarr

According to  appendix \ref{appendix 2}, equ. (\ref{raw scalar field equ}) 
the  scale covariant Euler-Lagrange equation is, for $D_{\phi}$ spacelike, 
 \[ 2 L_{HW}+4 L_{V4}-2 L_{\phi\,3} +\alpha \xi^2 \phi  \square\phi  - 2 (\xi \phi)^2 (\eta^{-1} \phi)^{-1}\,  \square_M \omega = 0  \,  .
\]

For $\alpha=6$ (and spacelike gradient $D\phi$ of the scalar field) 
 subtraction of the trace   of the Einstein equation simplifies the equation to:  
\beq  \square_M\, \omega= \frac{1}{2} ( \xi \phi)^{-2}  (\eta^{-1} \phi)\, \left(- \epsilon_{sig} \, tr\, T^{(m)}- 3 L_{\phi\,3} \right)  \,   \label{scalar field equ general}
\eeq 
  Both sides are of weight $-3$,  $tr\, T^{(m)}$ denotes the trace of the matter tensor. For $D\phi$ timelike or null and the choice (\ref{L_phi2 b}) for $L_{\phi\,2}$ the scalar field equation reduces to the potential condition
\beq L_{HW}+ 2  L_{V4}=0 \, .\label{pot condition}
\eeq 
In Einstein gauge (\ref{scalar field equ general}) becomes
\[   \square_M\, \omega \doteq 4 \pi G\, \tilde{a}_o\, tr\, T^{(m)} - ||\nabla \omega||^3 \, .
\]

We have to compose the scale covariant Milgrom operator from its Riemannian part and the scale connection component,
 $\square_M\, \omega= _g\hspace{-0.2em}\square_M\, \omega + _{\varphi}\hspace{-0.2em}\square_M\, \omega$,
 with the {\em covariant Milgrom operator} of Riemannian geometry defined by 
\beq _g\hspace{-0.12em}\square_M\, \omega = \epsilon_{sig} \, _g\hspace{-0.2em}\nabla_{\nu}(||\nabla \omega || \partial^{\nu} \omega) = \epsilon_{sig}\left(\partial_{\nu}||\nabla \omega || \, \partial^{\nu} \omega + || \nabla \omega || \, _g\hspace{-0.2em}\nabla_{\nu} \, \partial^{\nu} \omega   \right) \, . \label{covariant Milgrom op}
\eeq 
Because of $w(|| \nabla \omega|| \partial^{\nu}\omega)=-3$   we find 
\[ _{\varphi}\hspace{-0.2em}\square_M\, \omega = \epsilon_{sig}\left( _{\varphi}\hspace{-0.2em} \Gamma_{\nu \lambda}^{\nu}||\nabla \omega|| \partial^{\lambda}\omega  -3 \varphi_{\nu}||\nabla \omega||\partial^{\nu}\omega \right) =  \epsilon_{sig} \varphi_{\nu}||\nabla \omega||\partial^{\nu}\omega \, .
\] 
In Einstein gauge $_{\varphi}\hspace{-0.2em}\square_M\, \omega \doteq -||\nabla \omega||^3$.  
For a fluid with matter density $\rho_m$ and pressure $p_m$,  equation (\ref{scalar field equ general}) in Einstein gauge finally  simplifies to
\beq   _g\hspace{-0.12em}\square_M\, \omega \doteq  4 \pi G \, \tilde{a}_o\, tr\, T^{(m)} \doteq 4 \pi G \, \tilde{a}_o\, (\rho_m - 3p_m) \label{Milgrom equation}
\eeq 
By obvious reasons (\ref{Milgrom equation}) will be called the {\em covariant Milgrom equation}.

The Einstein equation (\ref{Einstein equation}) and the scalar field equation (\ref{scalar field equ general}), respectively (\ref{Milgrom equation}), constitute an interdependent system of differential equations.   We shall study it in the following section  under   simplifying conditions: a static weak field case and  a  cosmological limit.\footnote{In previous preprints of this paper the simplicity of (\ref{Milgrom equation}) could not be  achieved because no $L_{\phi\,2}$ term ($\alpha=6$) was included.}  
Before we do so, we want to  point out that the Schwarzschild-de Sitter solution is a special (point symmetric) vacuum solution  of  (\ref{Einstein equation}), (\ref{scalar field equ general})  with a trivial scalar field (constant in Riemann gauge).

\subsection{\small Schwarzschild-de Sitter solution \label{subsection Schwarzschild}}
Our first example deals with a Weyl geometrically degenerate case with  
Riemann gauge $(g, \varphi \doteq 0)$ identical to Einstein (scalar field gauge), $\phi \doteq \phi_o = const$. Here $g$ denotes the 
Schwarzschild-de Sitter metric of signature $(-+++)$:
\beq ds^2 = - (1-\frac{2M}{r} -\kappa\, r^2)dt^2 + (1-\frac{2M}{r} -\kappa\, r^2)^{-1}dr^2 + r^2(dx_2^2 + \sin^2 x_2\, dx_3)^2 \, 
\eeq
Then  $\square_M \, \omega =0,$ and (\ref{Milgrom equation}) is trivially satisfied in the vacuum. 

The Ricci and scalar cuvatures are   $Ric =3 \kappa\,g$, $R\doteq  12 \kappa$. We calculate in scalar field gauge, while  suppressing the dot of $\doteq$ here. The l.h.s. of our Einstein equation is familiar,
\[ Ric - \frac{R}{2}g = - 3\kappa\, g \, .
\]
In  vacuum  the r.h.s. of the Einstein equation  (\ref{ThetaI}, \ref{ThetaII}) simplifies to the quartic term (``cosmological constant'')  of the scalar field potential (\ref{cosmological constant}):
\[ \Theta = \Theta^{(I)}=-\frac{\lambda}{4}\xi^{-2} \phi_o^2 \, g = -\frac{\lambda}{4}\beta^2 \tilde{a_o}^2 \, g  \, 
\]
where $\beta$ denotes the ratio $\beta= \eta \,\xi^{-1}$ which, according to (\ref{geometrical mean}) is no large number. Then  (\ref{Einstein equation}) is satisfied for 
\[ 3 \kappa = \frac{\lambda}{4} \beta^2 \, \tilde{a_o}^2
\]
Below we shall find $\tilde{a}_o \approx \frac{a_o}{16} \approx 10^{-2}H$ (\ref{a_o-tilde}). With reasonable choices for $\beta \approx 100$ and, e.g.,  $\kappa =2H^2$ the equation is satisfied, for $\frac{\lambda}{4} \approx 6$.

Although this is a degenerate solution of the W-ST dynamical equations, it is important as a  non-homogeneous  {\em point symmetric} {\em vacuum solution} with $\nabla \omega =0$ (respectively with negligible gradient $\nabla \omega \approx 0$).  The deviation from the ordinary Schwarzschild equation is only by cosmologically small terms. It thus has the central symmetric point mass solution of the Newton theory as  its classical limit. In the next section we see that  another classical limit arises  as soon as we give up the degeneration condition Einstein gauge = Riemann gauge and we are far away from the source.

\newpage
\section{\small MOND approximation \label{section MOND approx}}
\subsection{\small  Modified Poisson equation for $\omega$  \label{subsection approx conditions}}

In the following we   assume  a  weak field constellation in which the Newton approximation of the Einstein equation is justified {\em even in the presence of a}  {\em scalar field $\omega$} with purely spacelike variability (signature choice $(-+++)$ of the metric,  $\epsilon_{sig}= +1$). This implies a 
\begin{itemize}
\item[($\ast$)] condition of {\em small} acceleration $a_{\varphi}= -\nabla \omega$, which has  to be specified in the particular cases studied.
\end{itemize}

In the case of exclusively  spacelike variability  the d'Alembert operator reduces (after sign change) to  the Laplacian,   $- _g\hspace{-0.1em}\square \, \, \omega =\Delta \omega$, and the Milgrom operator turns into the known form in spacelike coordinates,
\beq \square_M \omega \approx \nabla \cdot (| \nabla \omega | \nabla \omega) 
\eeq   
(``$\cdot $'' the Euclidean scalar product). For pressure-less matter the Milgrom equation (\ref{Milgrom equation}) acquires the familiar form of the non-linear Poisson equation\footnote{ \cite{Bekenstein/Milgrom:1984,Bekenstein:2004}.}
\beq \nabla \cdot (| \nabla \omega | \nabla \omega)  \approx 4 \pi G \; \tilde{a}_o\, \rho_m  \, .\label{non-linear Poisson equ}
\eeq 
We call this the {\em MOND approximation} of W-ST gravity. For the following it is important that  only  the trace of the {\em matter} energy momentum tensor, not of the scalar field, appears on the r.h.s. of (\ref{non-linear Poisson equ}).

Remember that $\omega $ is  the potential of an additional, not of the total, acceleration, and on the r.h.s of  (\ref{non-linear Poisson equ}) we still have the constant $\tilde{a}_o$ rather than $a_o$.  It will become clear from the growth behaviour of centrally symmetric solutions for $\omega$ that we cannot expect the conditions of a W-ST MOND approximation being satisfied in a full spherical neighbourhood of a centrally symmetric mass  concentration.   $\nabla \omega$ has to be   small enough for  the  energy tensor of the scalar field  to be such that its  approximative representation  in the Newton approximation leads to an acceptable approximation of the Einstein equation (sections \ref{subsection scalar field energy}, \ref{subsection a-add}).
This shows that the MOND approximation may be useful in large distance of a central mass only (if at all). In a closer vicinity the Schwarzschild-de Sitter solution with $\nabla \phi \approx 0$ will be a better approximation (section \ref{subsection Schwarzschild}) and with it, the Newton approximation, as long as relativistic effects can be neglected. 

A simple evaluation shows that for the Newton acceleration $a_N$, 
\beq \nabla^2 \Phi_N=4\pi G\, \rho_m \, \qquad a_N=- \nabla \Phi_N ,\label{a-N Phi-N}
\eeq
the solution of (\ref{non-linear Poisson equ}) is given by $\nabla \omega = - a_{\varphi}$ with
\beq   a_{\varphi} = \sqrt{\frac{\tilde{a}_o}{|a_N |}}\, a_N
=     \sqrt{\tilde{a}_o |a_N |}\, \frac{a_N}{| a_N|}   \, , \label{solution non-linear Poisson equ}
\eeq 
where $|a_N|$ is the vector norm in the Euclidean approximation.

This is a great relief: The solution of the non-linear Poisson equation is much simpler  than one might expect: At first the linear Poisson equation of the Newton theory is to be solved; then an algebraic transformation of type (\ref{solution non-linear Poisson equ}) leads to the solution of the non-linear partial differential equation (\ref{non-linear Poisson equ}).\footnote{In the terminology of the MOND community:  the MOND approximation of W-St leads to  a {\em QMOND model} \cite[pp. 46ff.]{Famaey/McGaugh:MOND}.}

For a point-like mass source $M$ at the origin of spatial  coordinates $y=(y_1,y_2,y_3) $,  the r.h.s  becomes $- \epsilon_{sig}  4 \pi G\;  tr\, T^{(m)} =  4\pi G\;  M \delta(y) $ . Considering an Euclidean approximation  for   $g_{\mu\nu}\dot{\approx} \,\eta_{\mu\nu} $, the corresponding  solution  is
\beq \omega \approx  \sqrt{GM \tilde{a}_o} \ln |y| \label{MOND solution} .\eeq
The Weyl geometric additional acceleration is
\beq
a_{\varphi} = - \nabla \omega  \,  \dot{\approx}   - \sqrt{GM \tilde{a}_o}\frac{y}{|y|^2}\, . \label{a-W point symmetric} \eeq
Its  form is the same as the  deep MOND acceleration of the usual MOND theory. Then
 \beq \Delta \omega  \;  \dot{\approx} \;  \frac{\sqrt{GM \tilde{a}_o}}{|y|^2} \, .  \label{Delta omega}
 \eeq  

The form of (\ref{a-W point symmetric}) shows that the MOND approximation can be reliable  only  in large distances from the symmetry centre; for `small' radii ($\ast$) is no longer satisfied. We shall use the specification $y\geq 10^{-l}\sqrt{GM a_o^{-1}}$. Then $\nabla \omega \leq 4\cdot 10^l a_o$. With  $l=1$  we  are at least in the  region called {\em upper transitional regime} in app. 7.3.\footnote{``At least'', because further out we enter the MOND regime or even the deep MOND regime of app. 7.3.} 

\subsection{\small Side remark on the cosmological limiting case \label{subsection cosmology}}
We want to make a short observation with regard to the cosmological limit. 
If we use the idealizing assumption   of  homogeneous matter distribution, the invariant scalar field does not depend on the spacelike coordinates of $x=(x_o,x_1,x_2,x_3), \; x_o=t$, 
\[
\omega(x)=\omega(t)\, , \qquad \nabla \omega = (g^{oo}\omega', 0,0,0)\, .
\] 
We then have a timelike or zero gradient of the scalar field $\nabla \omega$, respectively $D\phi$;  
$\left\| \nabla \omega \right\|$ vanishes, and with it $L_{\phi\,3}$  (\ref{L_phi 3}) and $\square_M$. For $L_{\phi\,2}$ we have to consider both choices (\ref{L_phi2 a}, \ref{L_phi2 b}).

For (\ref{L_phi2 a}) with $\alpha=6$, the scalar field equation reduces to the trace of the vacuum Einstein equation. In other words, it is compatible with the Einstein equation only for $tr\, T^{(m)}=0$, in which case it is redundant. 
An inspection of the Friedmann equation (in the Weyl gravity framework) in Einstein gauge shows that the Riemannian component of the scalar curvature must be constant. A special solution is  given by the Einstein - de Sitter model with warp function $a(t)=2 \sqrt{\frac{\Lambda}{3}}e^{\pm \sqrt{\frac{\Lambda}{3}} t}$ and vanishing gradient of the scalar field; i.e., the scalar field reduces to the cosmological constant, and Einstein gauge = Riemann gauge.

For (\ref{L_phi2 b}) both,  $L_{\phi\,2}$ and  $L_{\phi\,3}$, vanish  in the cosmological case. The scalar field equation reduces to the potential condition (\ref{pot condition}), and the trace of the Einstein equation to
\beq
 _g\hspace{-0.2em}\square\, \omega = \frac{4\pi G}{3} tr\, T^{(m)} \, .\label{tr Einstein cosmological}
\eeq
For a Robertson-Walker  metric in Einstein gauge
\[ g = \epsilon_{sig}\,diag\,(-1,a^2\,g_{11},a^2\,g_{22},a^2\,g_{33})
\]
with warp function $a=a(t)>0$, and time independent standard metric of constant curvature on the spacelike slices, $\tilde{g}=diag(g_{11},g_{22},g_{33})$ with. e.g., $g_{11}=(1- \kappa r^2)^{-1}, g_{22}= r^2, g_{33}=r^2 \sin^2 \theta\; (r=x_1, \theta = x_2)$,  we find
\[ \sqrt{|g|}=a^3 \sqrt{ |\tilde{g}| }, \qquad \frac{\partial_o \sqrt{|g|}}{\sqrt{ |g|}}=3 \frac{a'}{a}\, .
\]
Therefore with (\ref{Beltrami-Laplace})
\[ _g\hspace{-0.2em}\square\, \omega = \omega''+3 \frac{a'}{a}\, .
\]
and 
(\ref{tr Einstein cosmological}) becomes
\beq  \omega''+3 \frac{a'}{a}\omega'  = \frac{4 \pi G}{3} \, tr\,T^{(m)}\,.
 \eeq

 For the vacuum case  this  condition is satisfied by  a simple time-homogen\-eous static solution of the vacuum  Einstein  equation (\ref{Einstein equation}). In Einstein gauge it has the underlying Riemannian geometry of an Einstein universe  and a {\em  non-vanishing Weylian scale connection} $\varphi=(H,0,0,0)$ which encodes the cosmological redshift. This implies $\omega=-H\,t, \omega''=0$ and $a'=0$.\footnote{\cite{Scholz:BerlinEnglish,Scholz:FoP}.} 

 \subsection{\small Scalar field energy density \label{subsection scalar field energy}}
 We now want to address the distribution of the  scalar field's energy density. We use the static weak field approximation (\ref{weak field}) in Einstein gauge near a mass center. Then 
 $\omega(x)$ depends only on the spacelike coordinates of $x= (x_o, \ldots x_3)$, which we characterize separately  by the 3-vector $y:= (y_1,y_2,y_3) = (x_1,x_2,x_3)$.
 The energy-momentum tensor of the scalar field 
$ T^{(\phi)} \doteq (8\pi G)^{-1}\,\Theta   $
 is given by (\ref{ThetaI}), (\ref{ThetaII}). 
  Because of $\partial_o \omega=0$ the second term of the energy density of $\Theta^{(II)}$ vanishes immediately, and the first term in the light of  (\ref{D nu phi}). 

 In  the (static) weak field case, the cosmological constant contribution $L_{V4}\, g_{oo} $   lies many orders of magnitude below  energy densities considered here and can be neglected.  With  $g_{oo} \,  \dot{\approx} \, \eta_{oo} =   -\epsilon_{sig} \,$ we find
\beqarr \Theta_{oo} = \Theta_{oo}^{(I)} \, &{\approx}& \,  -  \phi^{-2}   \square \, \phi^2   - (\xi \phi)^{-2}L_{\phi} \; , \nonumber \\
 &{\approx}&  -  \phi^{-2}   \square \, \phi^2   - \frac{2}{3}(\eta^{-1} \phi)^{-1} \left\| \nabla \omega \right\|^3 \; . \label{Theta_oo^in rough}
\eeqarr
With (\ref{d Alembert of phi^2})
 (appendix \ref{appendix 1}) we get
 \beq  
 \Theta_{oo} \, {\approx} \,  -  2 \, _g\hspace{-0.1em}\square \, \omega   - \frac{2}{3}(\eta^{-1} \phi)^{-1} \left\| \nabla \omega \right\|^3   \, . \label{Theta_oo^in}
 \eeq
In the MOND and (upper) transitional regimes with, say  $|  a_N | \leq 10^{2}a_o$,  gives
\[  
\frac{2}{3} \tilde{a}_o \left\|\nabla \omega \right\|^3 \leq 10\,a_o^4 \, .
\]
It is cosmologically small of order 4 and thus negligible; hence  
\[ \Theta_{oo} \approx - 2  \, _g\hspace{-0.1em}\square \, \omega = 2   \Delta \omega \, . \qquad 
\]
The energy density of the scalar field, $\rho_{\phi}$, in Einstein gauge  finally becomes
\beq \rho_{\phi} \;  \dot{\approx} \;  (8\pi G)^{-1}\, 2\, \Delta  \omega \, . \label{scalar field halo density}
\eeq

  \subsection{\small Additional Newton acceleration and determination of $\tilde{a}_o$ \label{subsection a-add}}
 In the Newtonian limit case the energy of the scalar field  (\ref{scalar field halo density}) contributes to the right hand side of the Poisson equation and leads to   additional terms $\Phi_{\phi}$ and $a_{\phi}$ of the total Newton potential $\Phi_{tot}$ and its acceleration $a_{tot}$
\beqarr
\Phi_{tot} &=& \Phi_{N}+\Phi_{\phi}\, ,  \qquad \; a_{tot} = a_N+a_{\phi}\, ,\label{total Newton potential}\\
 \nabla^2 \Phi_N &=& 4\pi G\, \rho_m \, , \qquad  
\nabla^2 \Phi_{\phi} = 4\pi G\, \rho_{\phi}\, , \label{additional Newton pot}
\eeqarr 
where 
$a_{\phi}=-\nabla \Phi_{\phi}$ and $a_N$ like in (\ref{a-N Phi-N}).
The Poisson equation for $\Phi_{\phi}$ and (\ref{scalar field halo density}) imply
\beq a_{\phi} =-\nabla  \Phi_{\phi} =- \nabla \omega + X = a_{\varphi} +X\, ,
\eeq 
with a vector field  $X$ such that $\nabla X = \epsilon_{sig}\Gamma^j_{jk} \partial^k\omega$.

In the central symmetric case (not necessarily with a  point-like mass, but with total mass $M(r)$ inside the radius $r$ such that $M'(r)=0$ at  $r = |y|\geq r_o$ for some $r_o$) (\ref{solution non-linear Poisson equ}) implies:
\beqarr a_{\varphi}(y) &=& - \frac{\sqrt{\tilde{a}_o \, G\,M(r)}}{r}\frac{y}{r}  \nonumber \\
\omega &=& \sqrt{\tilde{a}_o \, G\,M(r)} \ln r \\
\nabla^2 \omega &=& \frac{\sqrt{\tilde{a}_o \, G\,M(r)}}{r^2}  \label{nabla^2 punktsymmetrisch}
\eeqarr 

With an Euclidean metric $ds^2 = dr^2 + r^2 (d\theta^2 + \sin^2 \theta\, d\vartheta^2)$  in spherical coordinates  $(r,\theta, \vartheta)$,\footnote{$\Gamma_{11}^1 =0,\, \Gamma_{2 1}^2 = \Gamma_{3 1}^3 = r^{-1}$.}
\[  \epsilon_{sig}\Gamma^j_{jk} \partial^k\omega = \frac{2}{r} \frac{\sqrt{\tilde{a}_o \, G\,M(r)}}{r} = 2\, \nabla^2 \omega
\]
and 
\beq X = 2\, a_{\varphi}\, .
\eeq 

We finally get an additional acceleration (with regard to the Newton acceleration $a_N$ of $\rho_m$) 
\beq a_{add} = a_{\varphi} + a_{\phi} = a_{\varphi} + 3\,a_{\varphi} = 4\, a_{\varphi}\, , \label{a-add}
\eeq 
and the total acceleration
\[ a = a_N + a_{add} = a_N + 4\, a_{\varphi} 
\]  
With   (\ref{solution non-linear Poisson equ})
\beq a=  a_N \left(1+ \sqrt{\frac{16\, \tilde{a}_o}{|a_N|}} \right) \label{total a rough}
\eeq 
Taking (\ref{a-W point symmetric}) into account, the total correction of the original Newton dynamics of a point-like (or point symmetric) source becomes 
\beq a_{add} = 4\, a_{\varphi} \approx  - 4 \sqrt{GM \tilde{a}_o}\,  \frac{y}{r^2}\, . \label{a_add final}
\eeq 

Now we  can specify the value of our $\tilde{a}_o$ for which our model gives a total additional acceleration which  {\em in the  deep MOND} domain agrees with the acceleration of Milgrom's MOND approach:
\beq \tilde{a}_o = \frac{a_o}{16} \approx \frac{H}{100} \approx 8\cdot10^{-31}\, cm \leftrightarrow 2\cdot 10^{-20}\, s^{-1}\, . \label{a_o-tilde}
\eeq 
Then (\ref{a_add final}) turns into
\beq a_{add} \approx -  \sqrt{GM {a}_o}\, \frac{y}{r^2} \, ,
\eeq 
with  the usual MOND acceleration $a_o \approx \frac{H}{6}[c]$, and   (\ref{total a rough}) becomes
\beq a= a_N \left(1+ \sqrt{\frac{a_o}{| a_N |}} \right) \,, \qquad a_{add} = \sqrt{a_o | a_N |}\frac{a_N}{| a_N |}\, . \label {a-add final}
\eeq

The norm of the complete (centrally oriented) radial acceleration in the MOND (and the transitional) regime about a point mass $M$, or in the case of a point symmetric mass distribution,  is given by (norm signs here omitted)
\beq 
a = a_N + a_{add} \approx   \frac{GM}{r^2} +  \frac{\sqrt{GM {a}_o}}{r} \, , \label{MOND acceleration}
\eeq
and the density of the scalar field halo   (\ref{scalar field halo density}) by
\beq \rho_{\phi}(r)  =  \frac{3}{4}(4 \pi G)^{-1}\,\frac{ \sqrt{GM{a}_o}}{r^2}  \,  .
\label{halo density}
\eeq 
In the case of a points symmetric mass distribution, $M$ has to be read as $M(r)$.
We resume: 
In the domain where the MOND approximation is reliable,   the acceleration correction to Newton gravity implied by the W-ST approach with cubic kinetical Lagrangian (for $\phi$) consists  simply in an {\em additive term} equal to the {\em deep MOND acceleration} of the usual MOND approach.

  \section{\small Comparison with other MOND models \label{subsection comparison}}
	\subsection{\small Transition function \label{subsection transition fct}}
We can now compare our approach with  other MOND models. Simply adding a deep MOND term to the Newton acceleration of a point  mass, like in (\ref{MOND acceleration}),  is unusual.  It is clear that such an approach does not lead to  acceptable results in the 'lower' transitional regime with, say, $a_N> 100\, a_o$ (app. \ref{appendix 3}).

M. Milgrom rather considered a multiplicative relation between the MOND acceleration $a$ and the Newton acceleration $a_N$ by a kind of `dielectric analogy':
\beq
a_N = \mu(\frac{a}{a_o}) \, a \; , \qquad \mbox{with} \quad \mu(x)\longrightarrow  \left\{{ 1 \quad  \;  \mbox{for}\; x \to \infty} \atop { x \quad \; \; \mbox{for} \; x\to 0 \; ,} \right. \label{mu function}
\eeq 
or the other way round
\beq
a = \nu (\frac{a_N}{a_o}) \, a_N \; ,  \qquad \mbox{with} \quad \nu(y) \longrightarrow 
\left\{{ 1 \hspace{2em} \mbox{for} \; y \to \infty} 
\atop {  y^{-\frac{1}{2}} \; \;   \mbox{for} \; y\to 0 \; . } \right. \label{nu function}
\eeq  
Here $\mu(x)\to x$ means $\mu(x)-x = \mathcal{O}(x)$, i.e. $\frac{\mu(x)-x}{x}$ remains bounded for $x \to 0$.
From this point of view our acceleration (\ref{MOND acceleration}) is specified by
\beq \mu_w(x) = 1+ \frac{1-\sqrt{1+4x}}{2x} \qquad \mbox{and}
\quad 
\nu_w(y) = 1+ y^{-\frac{1}{2}} \; . \label{our mu}
\eeq
One has to keep in mind that our transition functions $\mu, \nu$ are only reliable  in the MOND  and the upper transitional regimes  (section \ref{subsection approx conditions}).

This embedding into the MOND family shows that the so-called ``Kepler laws of galaxy dynamics'' hold  for our Weyl geometric scalar tensor (W-ST)  model like for all others in the family \cite[sec. 5]{Famaey/McGaugh:MOND}. 
But here, different from most  other family members, the MOND approximation results from a conceptually (with regard to space-time structure) and physically attractive (comparatively simple Lagrangian) {\em general relativistic ``mother'' theory}. Regarding the criteria of naturality and simplicity it may seem  superior to the better known relativistic MOND theories  TeVeS and Einstein aether theory.

 \subsection{\small Scalar field mass and phantom mass \label{subsection scalar field mass}}
It remains  to see how the Weyl geometric MOND model  compares  with the better studied ones with regard to rotation curves of galaxies, cluster dynamics, and lensing properties. Here we can give only a general overview of such a comparison; a detailed empirical evaluation  remains a desideratum.

 Equations (\ref{a-add}, \ref{a-add final}) show that three quarters of the W-ST additive acceleration are due to the scalar field energy density, the {\em scalar field halo}. That is important because the latter expresses a {\em true energy density} on the right hand side of the Einstein equation (\ref{Einstein equation}) and the Newtonian Poisson equation as  its weak field, static limit. It is decisive for {\em lensing} effects of the additional acceleration. 
In W-ST we have to distinguish between the influence of the additional structure, scalar field and scale connection, on light rays and on (low velocity) trajectories of mass particles. Bending of light rays is influenced by the scalar field halo only, the acceleration of  massive particles with velocities far below $c$ by the  the scalar field halo {\em and} the scale connection. 

In the MOND literature the amount of a (hypothetical) mass which  in Newton dynamics would produce the same effects as the respective MOND correction $a_{add}$ is called {\em phantom mass} $M_{ph}$. In our case, phantom mass and scalar field mass $M_{\phi}$ differ:
\beq  M_{ph} = \frac{4}{3} M_{\phi}
\eeq 

  For any member of the MOND family the additional acceleration can be expressed  by the modified  transition function 
	\beq  \tilde{\nu}= \nu -1  \label{nu tilde}
	\eeq
	with $\nu$ like in (\ref{nu function})
\beq a_{add}= \tilde{\nu}\left(\frac{|a_N|}{a_o}\right)\, a_N \, .
\eeq 
As the potential $\Phi_{ph}$ attributed  to the the phantom mass density $\rho_{ph}$ satisfies $\nabla^2 \Phi_{ph} =  4 \pi G\, \rho_{ph}$ and $\nabla \Phi_{ph}= - a_{add}$, a short calculation shows that the {\em phantom mass/energy density} may be expressed as
\beq  \rho_{ph}=  \tilde{\nu}\left(\frac{|a_N|}{a_o}\right)\, \rho_m 
-(4\pi G\, a_o)^{-1} \tilde{\nu}'\left(\frac{|a_N|}{a_o}\right)\, (\nabla| a_N |) \cdot a_N \label{rho-ph}
\eeq
It consists of a contribution proportional to $\rho_m$ with factor $\tilde{\nu}$, which dominates in regions of ordinary matter, and a term derived from the gradient of $|a_N|$ dominating in the ``vacuum'' (where however scalar field energy is present).
For the Weyl geometric model with $\tilde{\nu}_w(y)=y^{-\frac{1}{2}}, \, \tilde{\nu}_w'(y)=-\frac{1}{2}y^{-\frac{3}{2}}$ this implies:
\beqarr  \rho_{ph-w}&=& \left(\frac{a_o}{|a_N|}\right)^{\frac{1}{2}}\, \left( \rho_m + (8\pi G)^{-1}\, \nabla(| a_N |)\cdot \frac{a_N}{| a_N |} \right) \\
\rho_{\phi} &=& \frac{3}{4} \rho_{ph-w} \label{rho-ph Weyl}
\eeqarr
(\ref{rho-ph Weyl}) is  another expression for (\ref{scalar field halo density}).
Of course the terminology of ``phantom energy'' is misleading for $\rho_{ph-w}$, because three quarters of it are due to the scalar field and thus real  rather than phantom. 

The total dynamical mass  $M_{dyn}$ constituted by  a classical mass component (mainly baryonic), here denoted by $M_{bar}$, and phantom mass differs from  the  lensing mass $M_{lens}$:		
\beqarr  M_{dyn} &=& M_{bar} + M_{ph} \qquad   \label{M-dyn Weyl}\\
M_{lens} &=& M_{bar} + M_{\phi} = M_{bar} + \frac{3}{4}M_{ph} \qquad \mbox{in W-ST} \label{lensing mass}
\eeqarr
	In our model the lensing mass is {\em smaller} than the dynamical mass. That looks like bad news for explaining lensing at clusters and microlensing at substructures. But we shall see that the transition function  compensates this effect, perhaps even  more.
	
		\subsection{\small A first comparison between TeVeS and W-ST \label{subsection comparison TeVeS}}
	In the TeVeS literature it is taken for granted    that its scalar and vector fields, the additional structures of TeVeS,  influence  light trajectories like  a  real mass source of the same amount as the phantom mass would do in Einstein gravity  \cite[secs. 4f.]{Zhao_ea:2006}.
Therefore the {\em dynamical mass} $M_{dyn}$ and the {\em lensing mass} $M_{lens}$  are  identical,\footnote{\cite{Mavromatos_ea:2009} seem to 
doubt the reliabilty of the MOND approximations in some of the TeVeS calculations in the literature. They develop their own relativistic theory of light bending.} 
		\beq M_{dyn} = M_{lens} = M_{bar} + M_{ph} \,  \qquad \mbox{in TeVeS.}
		\eeq
Because of the factor $\frac{3}{4}$ in our (\ref{lensing mass}), lensing effects seem to be stronger in TeVes than in W-ST.
	But this inference is not  conclusive.  Phantom mass calculations depend strongly on the choice of the transition functions $\mu, \nu, \tilde{\nu}$ in the respective MOND model or their TeVeS equivalents.

	 The W-ST transition function $\nu_w$, respectively $\tilde{\nu}_w$   (\ref{our mu}) is larger than the  $\nu$-functions usually used in MOND/TeVes:  \cite{Mavromatos_ea:2009,Zhao_ea:2006} consider
	\[ \nu_1(y)=\frac{1}{2}(1+\sqrt{1+ 4\, y^{-1}})\] and $\nu_o$ corresponding to (Bekenstein's) $\mu_o(x)={2x}(1+2+\sqrt{1+4x})^{-1}$.  In his cluster studies R. Sanders uses\footnote{\cite{Sanders:1999,Sanders:2003}.}
	\[\nu_2(y)= \sqrt{\frac{1}{2}(1+\sqrt{1+4y^{-2}})}\, .\]  
		
\begin{figure}	
		{\includegraphics*[scale=0.5]{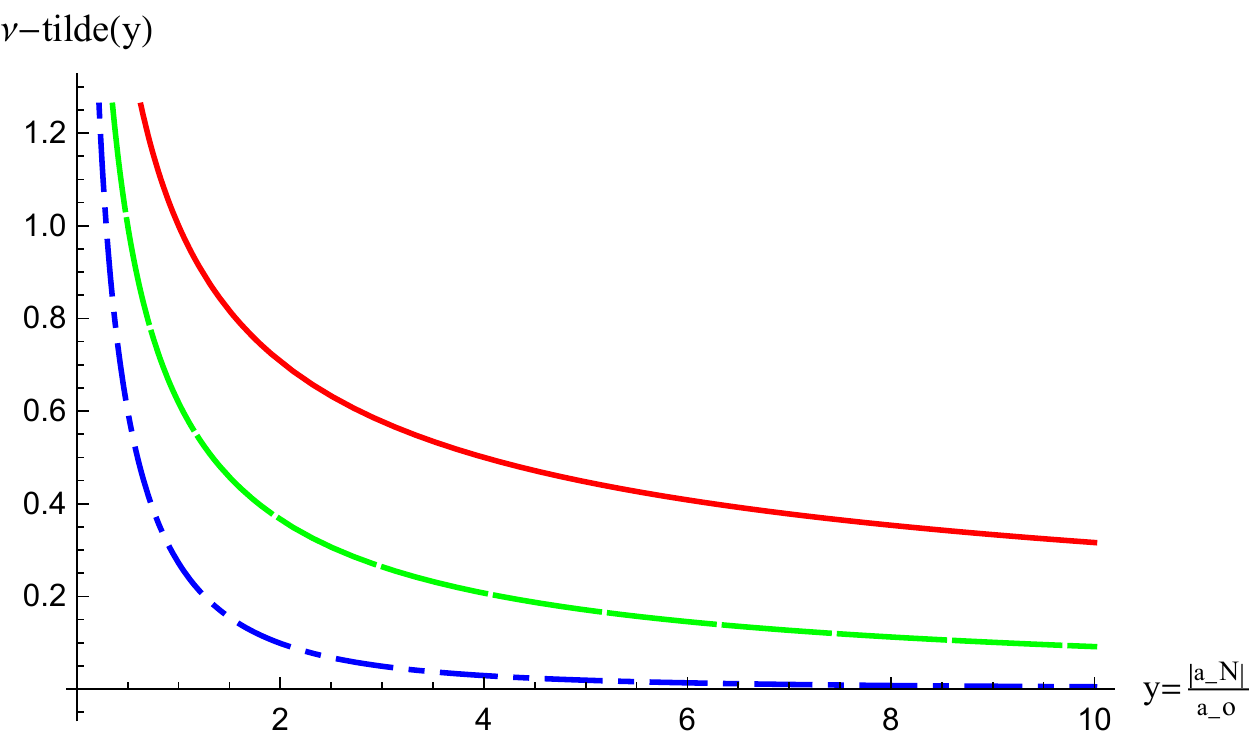}\includegraphics*[scale=0.5]{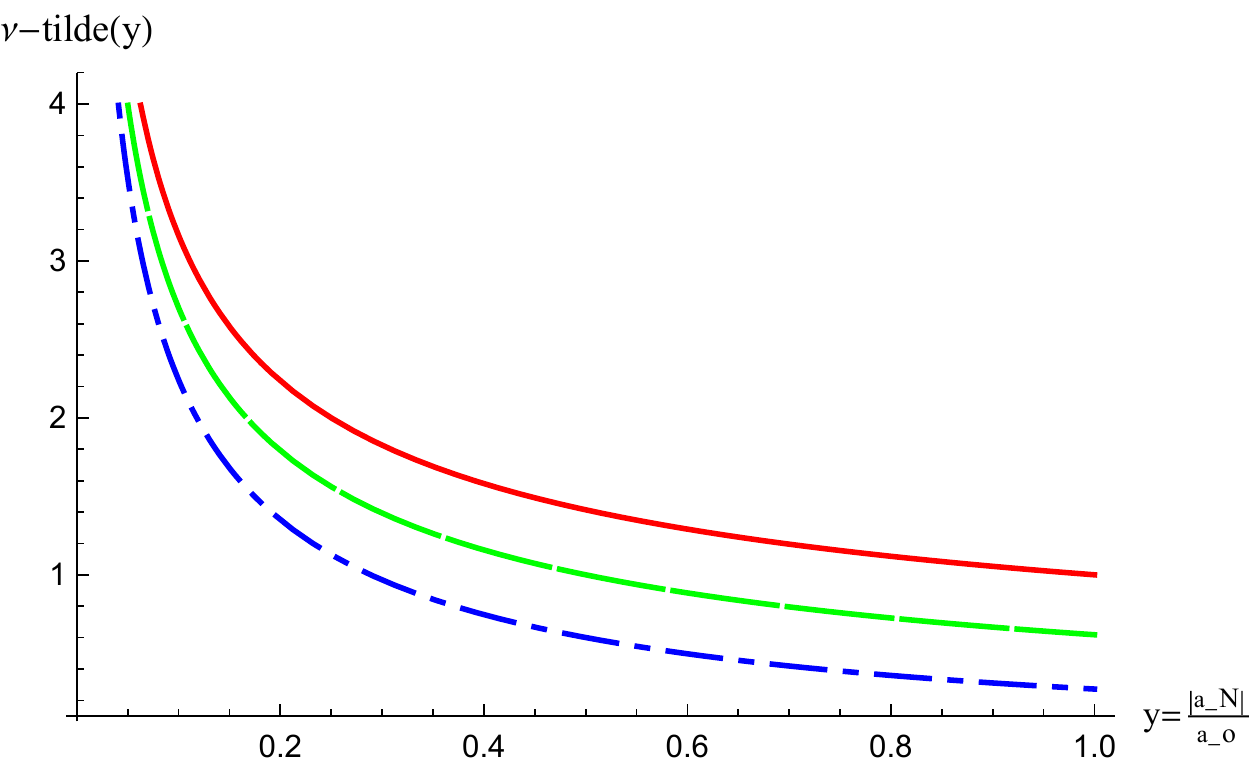}}
	\caption{\small Comparison of phantom halos   for Weyl model and usual MOND models \newline \hspace*{4em}  (for   $\tilde{\nu}$ see (\ref{nu tilde})). 
	Upper transition regime (left), MOND regime (right); \newline \hspace*{4em} 
	red/unbroken	$\tilde{\nu}_w(y)$: indicative of total phantom halo (scalar field and  \newline \hspace*{4em}  phantom) of Weyl model  (see (\ref{rho-ph})),    green/dashed $\tilde{\nu}_1(y)$, blue/double- \newline \hspace*{4em} dashed $\tilde{\nu}_2(y)$ for phantom halos  of accepted MOND models. }  
\vspace{2cm}

{\includegraphics*[scale=0.5]{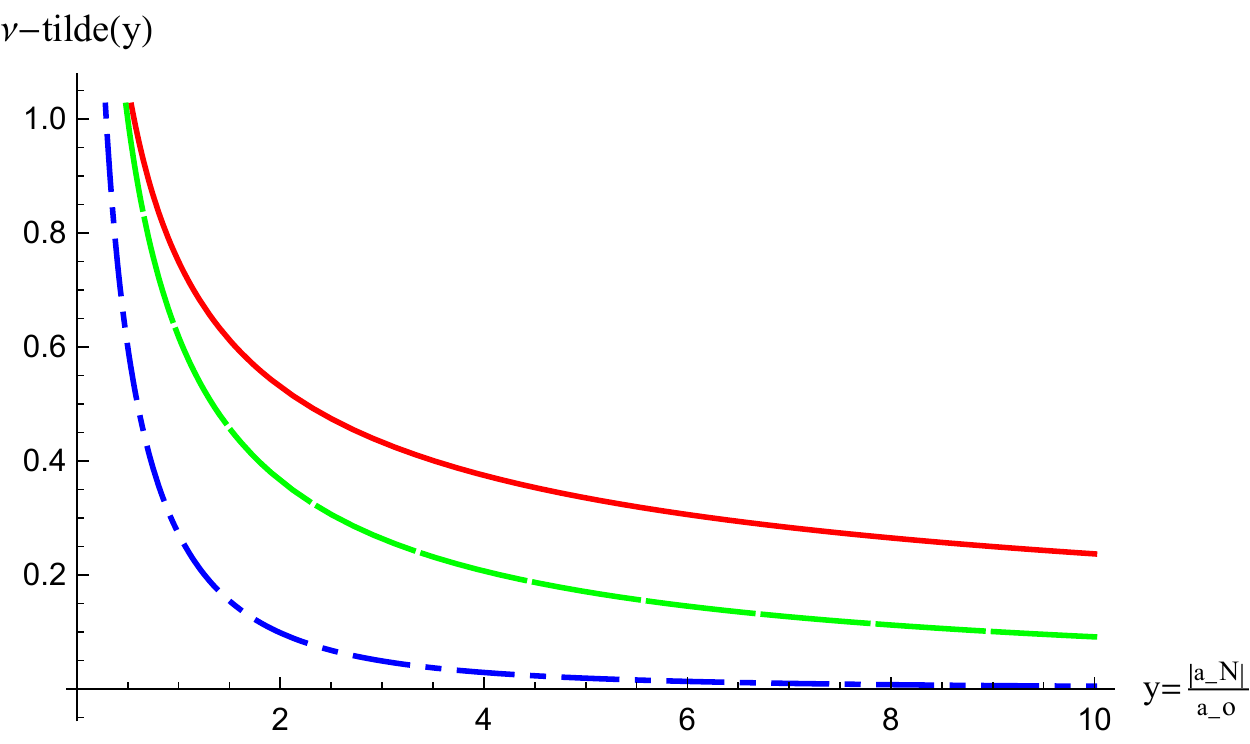} \includegraphics*[scale=0.5]{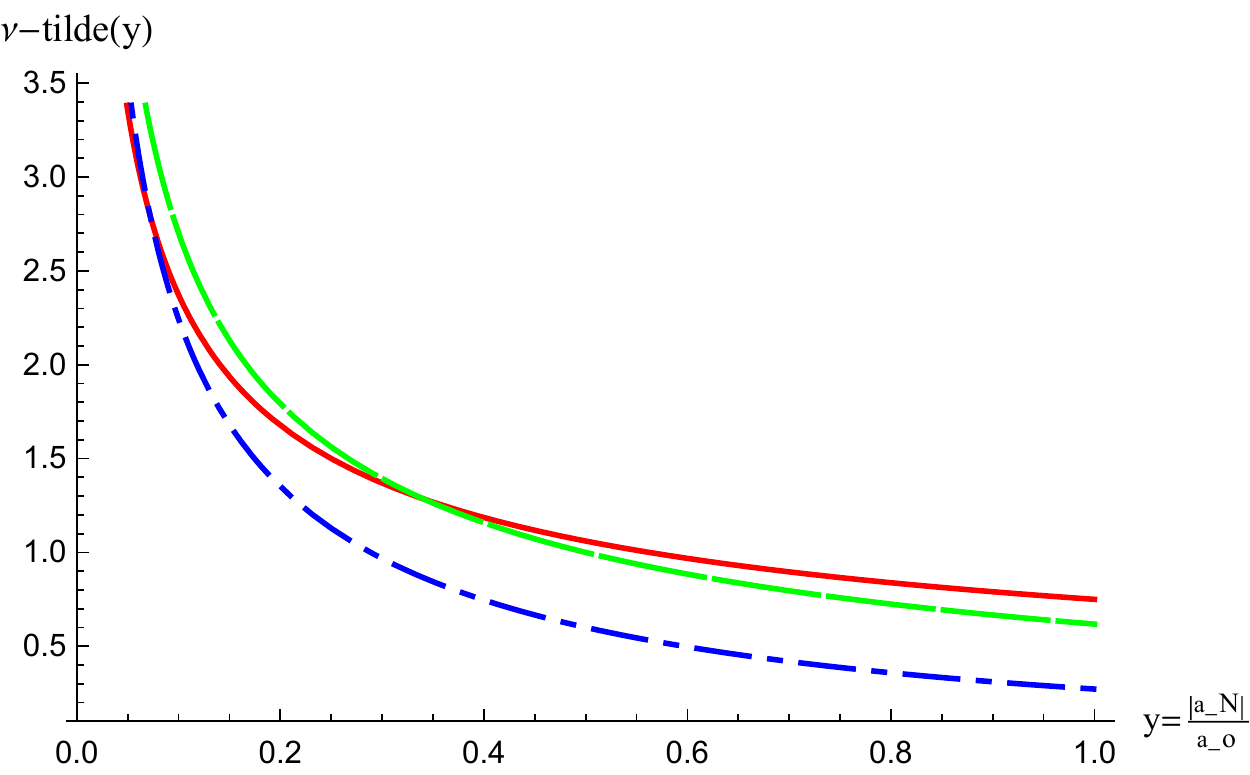} }
	\caption{\small    Comparison of scalar field halo for Weyl model with phantom halo 	of \newline \hspace*{4em} accepted MOND models  (for   $\tilde{\nu}$ see (\ref{nu tilde})). \newline \hspace*{4em} 
		Upper transition regime  (left) and MOND regime  (right); \newline
\hspace*{4.2em}   red/unbroken	$0.75\,\tilde{\nu}_w(y)$: indicative of scalar field halo Weyl model (\ref{rho-ph Weyl}),  \newline \hspace*{4em}  green/dashed $ \tilde{\nu}_1(y)$, blue/double-dashed $\tilde{\nu}_2(y)$ for phantom halos of  \newline \hspace*{4em}  widely used MOND models.}  
\end{figure}

The 		figures 1 and 2 below compare the Weyl geometric function $\frac{3}{4}\tilde{\nu}_w$ (red) governing the density of the  scalar field halo  with the typical MOND functions $\tilde{\nu}_1$ (green) of Mavromatos e.a. and $\tilde{\nu}_2$ (red) used by Sanders. The  $\tilde{\nu}$-term in (\ref{rho-ph}) dominates the respective phantom energy densities. Figure 1 shows  part of the  upper transition regime ($1\leq y \leq 10 $ with $y= \frac{|a_N|}{a_o}$) and figure 2  the beginning of the MOND regime ($0.1 \leq y \leq 1$). In the MOND regime $\rho_{\phi}$ is close to the phantom energy density of model $\nu_1$, but much higher than  $\nu_2$. In the transition regime the {\em W-ST scalar field halo is considerably denser than the phantom energy halo of both received MOND models}. The total phantom energy density of the Weyl approach, which is important for galaxy and cluster dynamics (\ref{M-dyn Weyl}),  comes out {\em even  higher} and surpasses  the phantom energies of the two other models in both domains (figures 3, 4).
	
These  considerations indicate that the missing mass problem for  clusters or galaxies, which is being discussed  for the MOND-TeVeS approach,\footnote{See, e.g., \cite{Famaey/McGaugh:MOND,Sanders:2003,Mavromatos_ea:2009,Zhao_ea:2006}.}
may  change its face in the  Weyl geometric approach.  
In the light of  the  comparison given in figures 3, 4, one might even hope that the mass discrepancy   
may dissolve under the present dynamical hypothesis.\footnote{A model for the halo of galaxy clusters, built on the MOND approximation (which here seems to be of heuristic value only) passes  a first empirical surprisingly well \cite{Scholz:2015Clusters}}
 But of course this is still far from clear; only  detailed empirical studies can show whether the Weyl geometric version of MOND-like weak gravity can really compete with, or even surpass, TeVeS and other relativistic MOND models.  In this respect, astronomers will have to speak the final word -- if there is any.


  \section{\small Discussion \label{section discussion}}
  Our assimilation of the original (R)AQUAL Lagrangian to Weyl geometric gravity has shown quite convincing properties. The Weyl geometric approach with its scale covariant expressions is conceptually  clearer than the 
	``2-metric-approach'' of the Jordan-Brans-Dicke framework in the AQUAL theory. Here Einstein gauge and Riemann gauge, or any other gauge, are mathematically equivalent. Which one seems best depends   on the specific problem context.
{\em Einstein gauge} gives the most  immediate expression to measured quantities; in this sense  it may  be considered as the {\em chronometric gauge}. But it would be misleading to call it {\em the} ``physical gauge''. The affine connection, and with it the {gravito-inertial structure} is most simply expressed in {\em Riemann gauge}. Whoever thinks of free fall as being governed by a Levi-Civita connection in the Riemannian sense, may just as well argue for Riemann gauge  as ``physical''.\footnote{This is reflected in the superiority of Riemann gauge for the variational procedures.}
 A coherent unification of the different aspects of spacetime structure is  made possible  by a consequently Weyl geometric perspective.  
  The additional degree of freedom (in comparison to Einstein  gravity) is related to the new dynamical variable $\omega$. It is regulated by the scalar field equation (\ref{Milgrom equation}). Because of (\ref{eq check varphi}) this equation can also be understood as a condition for the Weylian scale connection. In the degenerate case, $\omega= const$, the vacuum solution of a point mass source is the Schwarzschild-de Sitter solution with the classical Newtonian limit (section \ref{subsection Schwarzschild})
	
	A {\em first} dynamical consequence  of the Weyl geometric extension of Einstein gravity can be identified for low velocity trajectories in the weak field, static approximation  in Einstein gauge (the chronometric one). There the Weylian scale connection induces an additional acceleration to the usual  Newton approximation of Einstein theory (section \ref{subsection acceleration scale connection}). It has    the invariant scalar field $\omega$ as its potential (\ref{a_w}). 
	It seems quite natural to ask, whether this additional acceleration may be responsible for the anomalous effects of the MOND phenomenology; and if so, under which assumptions for the Lagrangian of the scalar field.

In the  {\em second step} we analyzed whether an adaptation of Be\-ken\-stein/Mil\-grom's non-quadratic Lagrange density  for the kinetic  term of the scalar field may help to  answer this question. Scale invariance gives a strong constraint for the form of the transition function; here it leads   to a  particularly simple, nearly unique, cubic form (\ref{cubic Lagrangian}). In an  approximation which allows to apply the Newton approximation of the Einstein equation,  the additional acceleration due to the  { scale connection} acquires a  {\em MOND-like} form (section \ref{section MOND approx}).
 So far our analysis is quite close to RAQUAL, the main differences being scale covariance and the fact that the Newton approximation of Einstein gravity remains a partial contribution of our MOND approximation  ((\ref{acceleration}), (\ref{a_R})).
  
  In a {\em third step} we have  analyzed the  energy density of the scalar field (sections \ref{subsection scalar field energy} and \ref{subsection a-add}) and found that it {\em modifies the total Newton potential} of the static weak field approximation  (\ref{scalar field halo density}), (\ref{total Newton potential}), (\ref{additional Newton pot}). That is a result of analyzing the r.h.s of the scale invariant Einstein equation; it needs no   additional stipulation. If compared with the original RAQUAL approach, this consequence of our approach changes the situation for {\em gravitational lensing} and for cluster dynamics considerably. 
  
  Given that the last mentioned problems (cluster dynamics and gravitational lensing) seem to have been most decisive for giving up the original RAQUAL approach,
one may ask why a similar observation has not been made already long since. The answer seems to reside in a widely spread conviction  that  scale covariant (or conformal) metrical approaches can never lead to a derivation of  gravitational lensing effects.  This conviction seems to have acquired the status of a kind of ``folk theorem''.\footnote{See, among others, \cite[pp. 146f.]{Sanders:DarkMatter}} 
 
This conviction has a true core, but it does not express the whole story. Like Diogenes who proved the possibility of motion to the Eleatic critics by walking, we have  shown that there {\em is} an alternative.   It is not difficult to see why it could work. The folk theorem has a premiss which often remains unstated. In the following quote it is stated explicitly:
 \begin{quote}
 ``\ldots {\em so long as the $\psi$ field} [corresponding to our $\omega$, E.S.] {\em contributes comparably little to the energy-momentum tensor}, it cannot affect light deflection \ldots''  \cite[p. 6, emph. E.S.]{Bekenstein:2004}. \end{quote}
  Why does this condition not apply to our Weyl geometric extension of essentially the same Lagrangian like in RAQUAL? 
  
  The answer can be read off from (\ref{ThetaI})  and (\ref{variation Hilbert-Weyl term}). The crucial difference in our energy-momentum tensor to the one often  used in  JBD-approaches,\footnote{Although some of the JBD literature does take account of the boundary terms of partial integration, e.g., \cite[pp. 40ff.]{Fujii/Maeda}.}
   comes from the  {\em boundary terms} arising during the  {\em variation of   the Hilbert action} to which the scalar field is non-minimally coupled.\footnote{See the literature in fn. \ref{fn Blago}.}
    Among  these terms, it is mainly $D_{\nu}D^{\nu}\phi^2$ which contributes essentially to the energy-momentum (\ref{Theta_oo^in rough}).
		  The sucessful adaptation of a cubic scalar field Lagrangian  to Weyl geometric gravity is a strong sign for the importance of the boundary terms. 
			
It is  too early to draw full consequences of this analysis at the moment. We still have to see whether the Weyl geometric approach proves to be of {\em empirical relevance} for extremely weak field domains at galaxy and perhaps even at galaxy cluster level, and whether a further analysis of domains, in which neither the MOND approximation nor the Schwarzschild-de Sitter approximation can be applied, sheds new {\em theoretical} light on strong field constellations. In the case of positive answers, or at least one with encouraging  result, we may conclude  that 	the energy density of the gravitational scalar field analyzed in our approach 	 is  {\em real} and not just a model artefact. First indications that the chances for a positive outcome of the empirical examination of our model are not bad are given section \ref{subsection comparison TeVeS}.

If so, we may interpret (\ref{ThetaI}), (\ref{ThetaII}), and (\ref{Theta_oo^in rough})  as expressions for the {\em    energy} of the (Weyl geometrically) enhanced {\em gravitational field}. Sceptics ought to remember that  a complete spacetime structure is given by the combination of a causal structure (mathematically a conformal structure), an inertio-gravitational  structure (projective path structure), and the scalar field specifying the remaining chronometric scaling degree of freedom,   mathematically  by the triple $( \mathfrak{c}, \nabla,\phi)$ (section \ref{subsection geodesics}).
{\em Gravitaty} is  a {\em complex  structure}, not just   one (vector, tensor, or connection) field. 

This insight may also  become important for  quantum gravity: In which sense could it be meaningful to quantize the  {\em basic  geometrical} features of spacetime, i.e. the conformal and affine structures $(\mathfrak{c},\nabla)$? It is well known that these structures do not carry intrinsic, covariant self energy, while  the scalar chronometric field does!  This speaks in favour of focussing the  quantization of gravity, at least in a first step, on the chronometric/scale degree of freedom $\phi$  and to analyze how the latter relates to the quantized standard model fields on general relativistic spacetime.


 \newpage
\section{\small Appendices}
\subsection{\small Scale invariant version of scalar field \label{appendix 1}} 
In Riemann gauge $(\tilde{g},0,\tilde{\phi})$ we write $\tilde{\phi}= e^{\omega}$ ($\omega$ stands here for the scale invariant form). 
By definition ${\omega}$ is not affected by  regauging, therefore
\beq D_{\nu}{\omega}= \partial_{\nu}{\omega}. \label{D_nu phi}\eeq
It is a {\em scale invariant} version of the {\em scalar field}.

Any scale gauge $(g,\varphi,\phi)$  arises from Riemann gauge, $g = \Omega^2 \tilde{g}$, for  some $\Omega$. Then 
\[ \varphi=-d\ln\Omega \leftrightarrow \Omega = e^{-\int \varphi}\,;\]
 here $\int \varphi$ is an abbreviated notation for integrating the 1-form $\varphi$ along any curve from a fixed initial point to the point $x$ of spacetime considered (underdetermination  only up to a point independent constant). We thus get
 \beqarr \tilde{\phi} &=& \Omega \phi\, , \nonumber\\
{\omega} &=& \ln{\tilde{\phi}}= \ln \phi - \int \varphi \, , \nonumber \\
\phi &=& \Omega^{-1} e^{{\omega}}= e^{{\omega}+\int \varphi}\,. \label{phi omega}
\eeqarr

In some of the recent literature  $\phi_{comp} :=  e^{\int \varphi}$ is considered on its own (with $\omega = 0$) \cite{Almeida/Pucheu:2014,Almeida/Pucheu_ea:2014}. It is a ``compensating field'' for the effects of a conformal transformation away from Riemann gauge.  Because of the gauge transformation for the scale connection it transforms with weight $w(\phi_{comp})=-1$ like $\phi$. But it does not essentially contribute to the dynamics besides giving it a scale covariant expression. Restricting   to $\phi_{comp}$  boils down to considering Einstein gravity in scale covariant form. The result  is a {\em dynamically trivial} Weyl geometric extension of Einstein gravity (and Riemannian geometry). 

If $(g,\varphi, \phi_o)$ denotes  a scalar field gauge, in  particular Einstein gauge $\phi_o \doteq \xi^{-1} E_{pl}$,  we have $\phi_o = \Omega^{-1}\tilde{\phi}$ with $\Omega= \phi_o^{-1}
e^{{\omega}}= \xi E_{pl}^{-1}e^{{\omega}}$; thus
$ \varphi \doteq - d \ln \Omega \doteq - d{\omega} $ and 
\beq
\varphi_{\nu} \doteq - \partial_{\nu} {\omega}. \label{omega potential}
\eeq
Thus $\omega$ has the formal properties of a potential for the scale connection $\varphi$ in {\em scalar field gauge} (and only in this gauge).

The scale covariant derivative of the scalar field in any gauge can be expressed as follows:
\beqarr D_{\nu}\phi &=& (\partial_{\nu}-\varphi_{\nu})\phi = \partial_{\nu}e^{{\omega}+\int \varphi} -\varphi_{\nu}\phi=(\partial_{\nu }{\omega} + \varphi_{\nu})\phi- \varphi_{\nu}\phi \nonumber \\
 &=& \phi\, \partial_{\nu}{\omega} = \phi\, D_{\nu}{\omega} \label{D nu phi}
\eeqarr
Similarly one derives 
 \beq D^{\nu} \phi^2 =  2 \phi^2 \partial^{\nu} \omega =   2 \phi^2 D^{\nu} \omega \, 
\eeq
and
\beqa D_{\nu}D^{\nu} \phi^2 &=& D_{\nu}(2\phi^2 D^{\nu}\omega) \\
&=& 2\phi^2 (D_{\nu}D^{\nu}\omega + 2 D_{\nu}\omega D^{\nu}\omega)\\
&=& 2\phi^2 (\nabla_{\nu}\partial^{\nu} \omega - 2 \varphi_{\nu}\partial^{\nu}\omega + 2 \partial_{\nu} \omega \partial^{\nu} \omega )
\eeqa
Because of $ _{\varphi}\hspace{-0.15em}\Gamma_{\nu \mu}^{\lambda}=\delta_{\nu}^{\lambda} \varphi_{\mu}+ \delta_{\mu}^{\lambda} \varphi_{\nu} - g_{\nu \mu}\varphi^{\lambda}$ 
 we find 
\beqa \nabla_{\hspace{-0.15em}\nu} \, \partial^{\nu} \omega &=& \, _g\hspace{-0.15em}\nabla_{\hspace{-0.2em}\nu}\,\partial^{\nu} \omega + _{\varphi \hspace{-0.2em}}\Gamma_{\nu \mu}^{\nu} \partial^{\mu}\omega \\
   &=& \, _g\hspace{-0.15em}\nabla_{\hspace{-0.2em}\nu}\,\partial^{\nu} \omega + 4 \varphi_{\nu}\partial^{\nu} \omega
\eeqa
In scalar field gauge $\varphi \doteq - \partial_{\nu} \omega$ and thus
\beq  D_{\nu}D^{\nu} \phi^2 \doteq 2 \phi^2\,  \, _g\hspace{-0.15em}\nabla_{\hspace{-0.2em}\nu}\,\partial^{\nu} \omega  \, . \label{D^2 partial omega}
\eeq 
In terms of the signature normalized Beltrami-d'Alembert operators (\ref{d Alembert}) 
\beq \square\, \phi^2 \doteq 2\phi^2 \, _g\hspace{-0.1em}\square\, \omega \, . \label{d Alembert of phi^2}
\eeq


\subsection{\small Derivation of the scalar field equation \label{appendix 2}}
We use scale covariant variation $\delta \omega$ with regard to the scale invariant scalar field $\omega$ as dynamical variable,  observing that $D_{\nu} \omega= \partial_{\nu} \omega$.\footnote{Equivalently, variation with regard  to $\phi$ could be taken.} 
We calculate the variation in Riemann gauge (then $R$ contains no $\varphi$-terms). Because of scale invariance of  the Lagrangian, the result translates straight forward to any gauge. 
The Lagrange density (\ref{L_phi}) is constructed   using  scale covariant derivatives $D_{\nu}$. The appropriate   Euler-Lagrange equation is  $\frac{\delta L_{\phi}}{\delta \phi}=\frac{\partial L_{\phi}}{\partial \phi} - D_{\nu}\frac{\partial L_{\phi}}{\partial(\partial_{\nu}\phi)}$ \cite[p. 526]{Frankel:Geometry}.

Using (\ref{partial omega of phi}) we get:
\beqa
\frac{\delta  {L}_{HW}}{\delta \omega} &=& \frac{\partial  {L}_{HW}}{\partial \omega} = \epsilon_{sig}\phi^2 R = 2  {L}_{HW}\, \\
\frac{\delta  {L}_{V4}}{\delta \omega} &=&\frac{\partial  {L}_{V4}}{\partial \omega} = 4  {L}_{V4}\, \\
\frac{\delta  {L}_{\phi\,2}}{\delta \omega} &=& \frac{\partial  {L}_{\phi\,2}}{\partial \omega} = 2{L}_{\phi\,2}\, , \quad \frac{\delta  {L}_{\phi\,3}}{\delta \omega} = \frac{\partial  {L}_{\phi\,3}}{\partial \omega} = {L}_{\phi\,3}
\eeqa
Moreover,
\[ 
\frac{\partial {L}_{\phi\,2}}{\partial (\partial_{\nu}\omega)} = \epsilon_{sig}\alpha (\xi \phi)^2 \partial^{\nu}\omega \, ,
\] 
and with (\ref{double partial omega}) 
\[ 
\frac{\partial {L}_{\phi\,3}}{\partial (\partial_{\nu}\omega)} =  2 \epsilon_{sig} \, \xi^2 \eta \,\phi \, \left\| \nabla \omega \right\|\partial^{\nu}\omega \, 
\]
for $\nabla \omega$ (respectively $D\phi$) spacelike (otherwise zero).
Because of
\beqa D_{\nu}\frac{\partial  {L}_{\phi\, 3}}{\partial (\partial_{\nu} \omega)}&=&  2 \epsilon_{sig} \, \xi^2 \eta \, D_{\nu} \left( \phi \,\left\| \nabla \omega \right\|\partial^{\nu}\omega  \right)\, \\
&=& 3 L_{\phi\, 3} + 2\xi^2 \eta\,  \phi \,D_{\nu}\left( \epsilon_{sig} \left\| \nabla \omega \right\| \partial^{\nu} \omega \right) \\
&=& 3 L_{\phi} + 2\xi^2 \eta \, \phi\; \square_M \omega  \qquad \mbox{(cf. (\ref{Milgrom operator}))}
\eeqa
and $D_{\nu}\frac{\partial  {L}_{\phi\, 2}}{\partial (\partial_{\nu} \omega)}= 2 L_{\phi\,2}+\epsilon_{sig}\alpha \xi^2 \phi D_{\nu}D^{\nu}\phi $,
this leads to the  ``raw'' scalar field equation
\beq 2 L_{HW}+4 L_{V4}-2 L_{\phi\,3} +\alpha \xi^2 \phi  \square\phi  - 2 (\xi \phi)^2 (\eta^{-1} \phi)^{-1}\,  \square_M \omega = 0  \,  \label{raw scalar field equ}
\eeq 
 for $D\phi$ spacelike. Otherwise, i.e. $D\phi$ causal, the $L_{\phi\,3}$ and $ \square_M \omega$ terms vanish; for the choice  (\ref{L_phi2 b}) also the  term in $\square\phi $.

On the other hand, tracing of  the Einstein equation (\ref{Einstein equation}) and multiplication by $\epsilon_{sig} (\xi \phi)^2$ leads to:
\beq 2 L_{HW} +  4 L_{V4} +  2\,(1-\frac{6}{\alpha}) L_{\phi\,2} +  L_{\phi\,3} + 6 \phi \xi^2 \square \phi +   \epsilon_{sig}\, tr\, T^{(m)} = 0 \label{tr Einstein equ}
\eeq

For $\alpha=6$ and spacelike $D\phi$ (respectively spacelike $\nabla \omega$) the subtraction of (\ref{tr Einstein equ}) from equ. (\ref{raw scalar field equ}) leads to the simplified scalar field equation  (\ref{scalar field equ general}),   of the main text: 
\[ 2 (\xi \phi)^{-2} (\eta^{-1} \phi)  \square_M\, \omega=   - \epsilon_{sig}\, tr\, T^{(m)}- 3\, L_{\phi\,3}   
\]
Without the $L_{\phi\,2}$ term and $\alpha=6$  additional 
 terms (in $\square \phi$ and proportional to $L_{\phi\,2}$) would appear.\footnote{They are called ``nuisance terms'' in the published version of this paper and in earlier preprints.} 
On shell of the Einstein equation  (\ref{Milgrom equation}) is equivalent to the raw scalar equation. For causal $D\phi$ and assuming $\alpha=6$, the scalar field equation is consistent with the Einstein equation only for $tr \, T^{(m)}=0$.

\subsection{\small MOND, deep MOND, and transition regimes \label{appendix 3}}
A  point  is  called to lie in the {\em MOND regime}, if the Newton acceleration  falls below $a_o$: $ a_{N} \leq a_o $ (here $a_N, \,  a_{add}$ denote the norm of the accelerations). In our approach  with additional acceleration  $a_{add}=\sqrt{a_N a_o}$ (\ref{a-add final}) this is equivalent to $a_{add}\geq a_N$.

If we agree to speak of {\em deep MOND} regime (dM), if the additional acceleration strongly dominates  the Newton acceleration, $a_{add}>> a_{N}$ in the sense of, say,  $a_{add}\geq 10\, a_{N}$ (or $a_{add}\geq 10^l\, a_{N}$), the dM condition is equivalent to $a_N\leq 10^{-2}\, a_o$ (respectively  $a_N\leq 10^{-2l}a_o$).

For $a_o \leq a_N \leq 100 a_o $ we speak of the {\em upper transition} regime from Newton to MOND. For the 'lower' transition regime with $a_N> 100 a_o$ the MOND approximation of W-ST loses its reliability (section \ref{subsection approx conditions}).

In short: we have dM for $\frac{a_N}{a_o}\leq 10^{-2}$, MOND regime for $\frac{a_N}{a_o}\in [0.01,1]$, and the upper transition regime if $\frac{a_N}{a_o}\in [1,100]$  (for $k=l=1$). 
For a central symmetric mass $M$ the MOND regime starts at the distance $r_o=\sqrt{GM a_o^{-1}}$, the transition regime at $10^{-1}\,r_o$,  dM  at  $10\,r_0$.


For stars with size of the sun,   $G M_{\astrosun} \sim 10^5 \, cm$, and with $a_o \sim \frac{H}{6} \sim 10^{-29} \, cm^{-1}$ we get $r_o \sim 10^{17}\, cm \sim 10^4\, AU \sim 10^{-1}\, pc$. For the mass of a galaxy with $M_{gal}\sim 10^{11}M_{\astrosun}$, idealized to spherical symmetry, the  MOND regime of the total galaxy begins  5 to 6 orders of magnitude higher, $r_o \sim 10\, kpc$, the deep MOND at the outskirts of the disk $R_1 \sim 100\, kpc$. Note that the stars constituting the galaxy have their own MOND and dM regimes at the lower scale. In our approach, their scalar field halos  contribute   to the total gravitational mass-energy of the galaxy and are crucial for microlensing effects.

\small
 \bibliographystyle{apsr}
  \bibliography{a_lit_hist,a_lit_mathsci}
\end{document}